\def\ni{\noindent}
\def\vs{\vskip.3cm}

\def\+{{(+)}}  \def\-{ {(-)} }   \def\0{ {(0)} }
\def\1{ {(1)} }  \def\2{ {(2)} }

\def\sq{Q\kern-6pt/}
\def\sQ{Q\kern-12pt\nearrow}
\documentclass[11pt,eqs]{article}

\usepackage{latexsym}
\usepackage{amssymb}

\def\be{\begin{equation}}             \def\ee{\end{equation}}
\def\ba{\begin{array}{rcl}}           \def\ea{\end{array}}
\def\beqa{\begin{eqnarray} }          \def\eeqa{\end{eqnarray} }
\def\beqalign{\begin{eqalign}}        \def\eeqalign{\end{eqalign}}
             
\def\bsubeq{\begin{subequations}}     \def\esubeq{\end{subequations}}
\def\bitem{\begin{itemize}}           \def\eitem{\end{itemize}}

\def\DJ{\leavevmode\setbox0=\hbox{D}\kern0pt
 \rlap{\kern.04em\raise.188\ht0\hbox{-}}D}
\def\dj{\leavevmode\setbox0=\hbox{d}\kern0pt
 \rlap{\kern.215em\raise.46\ht0\hbox{-}}d}

\newcommand{\bd}{\begin{displaymath}}
\newcommand{\ed}{\end{displaymath}}

\begin{document}

\title{ $D5$-brane type I superstring background fields in terms of type IIB ones by canonical method and T-duality approach
\thanks{Work supported in part by the Serbian Ministry of Science and
Technological Development, under contract No. 141036. Bojan
Nikoli\'c also acknowledge hospitality and useful discussions at
the Institute for Nuclear Research and Nuclear Energy in Sofia
(Bulgaria) during his visit as early stage researcher supported by
the FP6 Marie Curie Research Training Network "Forces-Universe"
MRTN-CT-2004-005104.}}
\author{B. Nikoli\'c \thanks{e-mail address: bnikolic@ipb.ac.rs} and B. Sazdovi\'c
\thanks{e-mail address: sazdovic@ipb.ac.rs}\\
       {\it Institute of Physics, 11001 Belgrade, P.O.Box 57, Serbia}}
\maketitle
\begin{abstract}

We consider type IIB superstring theory with embedded $D5$-brane
and choose boundary conditions which preserve half of the initial
supersymmetry. In the canonical approach that we use, boundary
conditions are treated as canonical constraints. The effective
theory, obtained from the initial one on the solution of boundary
conditions, has the form of the type I superstring theory with
embedded $D5$-brane. We obtain the expressions for $D5$-brane
background fields of type I theory in terms of the $D5$-brane
background fields of type IIB theory. We show that beside known
$\Omega$ even fields, they contain squares of $\Omega$ odd ones,
where $\Omega$ is world-sheet parity transformation,
$\Omega:\sigma\to -\sigma$. We relate
result of this paper and the results of \cite{BNBSPLB} using T-dualities along four directions orthogonal to $D5$-brane.

\end{abstract}
\vs

\ni {\it PACS number(s)\/}: 11.25.Uv, 11.25.-w, 04.20.Fy   \par

\section{Introduction}

It is a known fact that states of the type IIB superstring theory
even under world-sheet parity projection $\Omega:\sigma\to-\sigma$
correspond to the states of type I superstring theory
\cite{jopol}. More precisely, the states, which correspond to the
background fields: graviton $G_{\mu\nu}$ and dilaton $\Phi$ from
NS-NS sector, the sum of two same chirality gravitinos
$\psi^\alpha_{+\mu}=\psi^\alpha_\mu+\bar\psi^\alpha_\mu$ and
dilatinos $\lambda^\alpha_+=\lambda^\alpha+\bar\lambda^\alpha$
from NS-R sector and two rank antisymmetric tensor $A_{(2)}$ from
R-R sector, survive this projection. The states that are odd under
$\Omega$ transformation: antisymmetric tensor $B_{\mu\nu}$ from
NS-NS sector, difference of two gravitinos
$\psi^\alpha_{-\mu}=\psi^\alpha_\mu-\bar\psi^\alpha_{\mu}$ and
dilatinios $\lambda^\alpha_{-}=\lambda^\alpha -\bar\lambda^\alpha$
from NS-R sector, and scalar $A_{(0)}$ and four rank antisymmetric
tensor $A_{(4)}$ with self dual field strength from R-R sector,
are eliminated by above projection.

Here we will consider the propagation of the open string in the
background of type IIB theory. In order to clarify notation and
terminology we will distinguish two descriptions of the same
theory. We start with variables $x^\mu$, $\theta^\alpha$ and $\bar
\theta^{\alpha}$ and background fields $G_{\mu\nu}$, $B_{\mu\nu}$,
$\Psi^\alpha_\mu$, $\bar\Psi^\alpha_\mu$ and $F^{\alpha\beta}$
where the theory is described by equations of motion and boundary
conditions. We are able to solve boundary conditions and introduce
the effective theory defined only by equations of motion. It turns
out that this effective theory is again the string theory, but in
terms of effective coordinates and effective background fields. As
a consequence of the boundary conditions, effective theory is
$2\pi$ periodic and describes propagation of closed string in the
background of type I theory.

In Ref.\cite{BNBSPLB} we investigated the relation between type
IIB and type I superstring theories in pure spinor formulation
\cite{berko,susyNC}. It turned out that effective theory, obtained
from initial one on the solution of boundary conditions, is just
type I closed superstring theory. We improved known expressions
for type I background fields with terms bilinear in $\Omega$ odd
fields of type IIB superstring theory. In the present paper we
extend the result of Ref.\cite{BNBSPLB} embedding $Dp$-brane so
that string endpoints move along it. Let us note that the action,
which we used in Ref.\cite{BNBSPLB} and in the present paper, can
be obtained from general expression for type IIB superstring
action \cite{NPBref} requiring that all background fields are
constant and neglecting all nonquadratic terms. Consequently, all
results we obtain are valid up to quadratic level.

We want to have stable $Dp$-branes \cite{jopol} both in initial
and final (effective) theory. Electrical charge and charge of
magnetic dual brane in $D=10$ are given by the following
expressions, respectively
\begin{equation}
e_p=\int_{S^{8-p}} {}^\star F_{(8-p)}\, ,\quad \mu_{6-p}=\int_{S^{p+2}} F_{(p+2)}\, ,\nonumber
\end{equation}
where $F_{(p+2)}=d A_{(p+1)}$ is field strength and ${}^\star F_{(8-p)}$ is its Hodge
dual. The R-R sector of type IIB theory contains
gauge fields $A_{(0)}$, $A_{(2)}$ and $A_{(4)}$. In effective,
type I superstring theory, only two form gauge field, $A_{(2)}$,
exists. It couples electrically to $D1$-brane ($e_1$) and
magnetically to $D5$-brane ($\mu_5$). In order to work with stable
$Dp$-branes in both theories, we will embed $D5$-brane.

We choose Neumann boundary conditions for $x^i$ coordinates
($i=0,1,\dots,5$), and Dirichlet boundary conditions for the rest
ones $x^a$ ($a=6,\dots,9$). In this way we embed $D5$-brane in
type IIB and type I theories in $D=10$, and break the initial
symmetry $SO(1,9)$ to $SO(1,5)\times SO(4)$.

Any $D=10$ dimensional Majorana-Weyl spinor
$S^\alpha\,(\alpha=1,2,\dots,16)$ can be expressed in terms of two
$D5$-brane opposite chirality Weyl spinors, $S^{\alpha_1}$ and
$S^{\alpha_2}$ $(\alpha_1,\alpha_2=1,2,\dots,8)$
\cite{jopol,duf,grk}. According to this decomposition of spinors,
the ten dimensional bispinor $F^{\alpha\beta}$ can be expressed in
terms of 8 independent  $D5$-brane bispinors (Appendix B). It
turns out that R-R sector on $D5$-brane contains four scalars
$a_{(0)}$, four vectors $a_{(1)}$ and four two rank antisymmetric
tensors $a_{(2)}$ with self-dual field strengths. For fermionic
$D5$-brane coordinates we choose,
$(\theta^{\alpha_1}-\bar\theta^{\alpha_1})|_0^\pi=0$ and
$(\theta^{\alpha_2}+\bar\theta^{\alpha_2})|_0^\pi=0$, which
produces corresponding boundary conditions for canonically
conjugated momenta, $(\pi_{\alpha_1}-\bar\pi_{\alpha_1})|_0^\pi=0$
and $(\pi_{\alpha_2}+\bar\pi_{\alpha_2})|_0^\pi=0$.

In our approach boundary conditions are treated as canonical
constraints. It turns out that all constraints originating from
boundary conditions are of the second class. Solving the second
class constraints, we obtain effective theory, which is described
by the $\Omega$ even Lagrangian. We recognize effective theory as
type I theory with embedded $D5$-brane. Consequently, we obtain
the expressions for $D5$-brane background fields of type I
superstring theory in terms of corresponding ones of type IIB.
More precisely, $D5$-brane background fields of type I theory,
beside known term with $\Omega$ even fields, contain new term with
quadratic combinations of $\Omega$ odd $D5$-brane background
fields of type IIB theory. The quadratic parts of effective
background fields can be considered as supersymmetric
generalization of the open string metric $G^{eff}_{\mu\nu}$
obtained by Seiberg and Witten \cite{SW}.

The
expressions for $D5$-brane background fields of type I theory can be obtained directly from the expressions for
effective background fields of Ref.\cite{BNBSPLB} using T-dualities along $x^a$ directions \cite{JT}.

At the end we give some concluding remarks. Also there are four
appendices. In Appendix A we introduce representation of gamma
matrices adopted from first reference in \cite{jopol} and then
consider the spinors in ten and six dimensions and relations
between them. The Appendix B deals with bispinors in ten and six
dimensions. We showed that ten dimensional bispinor can be
expressed in terms of eight independent $D5$-brane bispinors.
Discussion about complex coordinates and their canonically
conjugated momenta is given in the Appendix C. The complete
consistency procedure for fermionic constraints is presented in
Appendix D.

\section{Embedding $D5$-brane in type IIB theory}
\setcounter{equation}{0}

In this section we will consider action for type IIB theory in
pure spinor formulation \cite{berko,NPBref}. This theory is
originally formulated using BRST charge and contains ghost fields.
As in \cite{susyNC}, we drop ghost terms and consider only ghost
independent part of the action. We will preserve the parts of
background fields which are nontrivial on the embedded $D5$-brane,
where the open string endpoints are attached.

Let us start with sigma model action for type IIB superstring of
Ref.\cite{NPBref}
\begin{equation}
S=S_0+V_{SG}\, .
\end{equation}
The action in a flat background
\begin{equation}
S_0=\int_\Sigma d^2\xi \left( \frac{\kappa}{2}\eta^{mn}\eta_{\mu\nu}\partial_m x^\mu \partial_n x^\nu-\pi_\alpha \partial_{-} \theta^\alpha+\partial_+ \bar\theta^\alpha \bar\pi_\alpha\right)\, ,
\end{equation}
is deformed by integrated form of the massless IIB supergravity
vertex operator
\begin{equation}
V_{SG}=\int_\Sigma d^2 \xi X^T_M A_{MN}\bar X_N\, .
\end{equation}
The world sheet $\Sigma$ is parameterized by $\xi^m=(\xi^0=\tau\,
,\xi^1=\sigma)$, and $D=10$ dimensional space-time is
parameterized by coordinates $x^\mu$ ($\mu=0,1,2,\dots,9$). The
fermionic part of superspace is spanned by same chirality
fermionic coordinates $\theta^\alpha$ and $\bar\theta^{\alpha}$,
while the variables $\pi_\alpha$ and $\bar \pi_{\alpha}$ are their
canonically conjugated momenta. These fermionic coordinates and
momenta are Majorana-Weyl spinors in $D=10$ dimensions. We use
notation: $\dot X\equiv\partial_\tau X$, $X'\equiv\partial_\sigma
X$ and $\partial_\pm=\partial_\tau\pm\partial_\sigma$, where $X$
is an arbitrary function of $\tau$ and $\sigma$.

For our purpose it is enough to consider the part of left and
right supersymmetric tensors
\begin{equation}
X_M=\left(\begin{array}{c}
\partial_+\theta^\alpha\\
\Pi_+^\mu\\
d_\alpha
\end{array}\right)\, ,\quad \bar X_N=\left(\begin{array}{c}
\partial_-\bar\theta^\alpha\\
\Pi_-^\mu\\
\bar d_\alpha
\end{array}\right)\, ,
\end{equation}
and matrix
\begin{equation}\label{eq:Amn}
A_{MN}=\left(\begin{array}{ccc}
A_{\alpha\beta} & A_{\alpha\nu} & E_\alpha{}^\beta\\
A_{\mu\beta} & A_{\mu\nu} & \bar E_\mu^\beta\\
E^\alpha{}_\beta & E^\alpha_\nu & P^{\alpha\beta}
\end{array}\right)\, .
\end{equation}
Matrix with superfields generally depends on $x^\mu$,
$\theta^\alpha$ and $\bar\theta^\alpha$.

The BRST invariance of vertex operator produces equations of
motion
\begin{equation}
\Gamma^{\alpha\beta}_{\mu\nu\rho\sigma\lambda}D_\alpha A_{\beta\gamma}=0\, ,\quad \Gamma^{\alpha\beta}_{\mu\nu\rho\sigma\lambda}\bar D_\alpha A_{\gamma\beta}=0\, .
\end{equation}
We will additionally require that all background fields in
(\ref{eq:Amn}) are constant and restrict the analysis only to the
quadratic terms. With these assumptions there exists simple
solution
\begin{equation}
\Pi_\pm^\mu\to \partial_{\pm} x^\mu\, ,\quad d_\alpha\to \pi_\alpha\, ,\quad \bar d_\alpha\to \bar\pi_\alpha\, ,
\end{equation}
and only nontrivial superfields take the form
\begin{equation}
A_{\mu\nu}=\kappa(\frac{1}{2}g_{\mu\nu}+B_{\mu\nu})\, ,\quad E^\alpha_\nu=-\Psi^\alpha_\nu\, ,\quad \bar E_\mu^\alpha=\bar\Psi_\mu^\alpha\, ,\quad P^{\alpha\beta}=\frac{1}{2\kappa}F^{\alpha\beta}\, ,
\end{equation}
where $g_{\mu\nu}$ is symmetric and $B_{\mu\nu}$ is antisymmetric
tensor. We adopt expressions to be in agreement with our
conventions.

Under imposed condition we obtain the vertex operator
\begin{eqnarray}
V_{SG}=\int_\Sigma d^2\xi \left[
\kappa(\frac{1}{2}g_{\mu\nu}+B_{\mu\nu})\partial_+x^\mu\partial_-x^\nu-\pi_\alpha
\Psi^\alpha_\mu \partial_-x^\mu+\partial_+ x^\mu
\bar\Psi^\alpha_\mu\bar\pi_\alpha+\frac{1}{2\kappa}\pi_\alpha
F^{\alpha\beta}\pi_\beta\right]\, .
\end{eqnarray}
Together with flat background action it produces
\begin{eqnarray}\label{eq:SB}
&{}&S=\kappa \int_\Sigma d^2\xi \left[
\frac{1}{2}\eta^{mn}G_{\mu\nu}+\varepsilon^{mn}
B_{\mu\nu}\right]\partial_m x^\mu \partial_n x^\nu
\\&+&\int_\Sigma d^2 \xi \left[ -\pi_\alpha
\partial_-(\theta^\alpha+\Psi^\alpha_\mu
x^\mu)+\partial_+(\bar\theta^{\alpha}+\bar \Psi^{\alpha}_\mu
x^\mu)\bar\pi_{\alpha}+\frac{1}{2\kappa}\pi_\alpha F^{\alpha
\beta}\bar \pi_{\beta}\right ]\, ,\nonumber
\end{eqnarray}
where $G_{\mu\nu}=\eta_{\mu\nu}+g_{\mu\nu}$.

Choosing Neumann boundary conditions for $x^i\, (i=0,1,\dots,5)$
and Dirichlet boundary conditions for orthogonal directions $x^a\,
(a=6,7,8,9)$, we embed $D5$-brane in $D=10$ dimensional
space-time. Orthogonality of these two sets of coordinates implies
$G_{ia}=0$. We assume that antisymmetric Neveu-Schwarz field
$B_{\mu\nu}$ is nontrivial only along $D5$-brane, $B_{\mu\nu}\to
B_{ij}$. In NS-R sector the nonzero components $\Psi^\alpha_i$ and
$\bar\Psi^{\alpha}_i$ can be expressed in terms of $D5$-brane Weyl
spinors $\Psi^{\alpha_1}_i$, $\Psi^{\alpha_2}_i$,
$\bar\Psi^{\alpha_1}_i$ and $\bar\Psi^{\alpha_2}_i$ (see Appendix
A). The rest ones $\Psi^\alpha_a$ and $\bar\Psi^\alpha_a$ are set
to zero. In the R-R sector we assume nonzero value of all
$D5$-brane bispinors (see Appendix B). Because we restricted our
analysis to quadratic terms, the part of the action describing the
free string oscillation in $x^a$ directions decouples from the
rest. Taking into account all these assumptions, the action gets
the form
\begin{eqnarray}\label{eq:SB1}
&S&=\kappa \int_\Sigma d^2\xi \left[
\frac{1}{2}\eta^{mn}G_{ij}+\varepsilon^{mn}
B_{ij}\right]\partial_m x^i \partial_n x^j\nonumber
\\&+&2\Re\left\lbrace \int_\Sigma d^2 \xi \left[  -\pi_{\alpha_1}
(\partial_\tau-\partial_\sigma)\left(
\theta^{\alpha_1}+\Psi^{\alpha_1}_i x^i\right)
+(\partial_\tau+\partial_\sigma)\left( \bar\theta^{\alpha_1}+\bar
\Psi^{\alpha_1}_i x^i\right) \bar\pi_{\alpha_1}\right]\right\rbrace  \nonumber
\\ &+&2\Re\left\lbrace \int_\Sigma d^2 \xi \left[  -\pi_{\alpha_2}
(\partial_\tau-\partial_\sigma)
\left(\theta^{\alpha_2}+\Psi^{\alpha_2}_i x^i\right)
+(\partial_\tau+\partial_\sigma) \left(\bar\theta^{\alpha_2}+\bar\Psi^{\alpha_2}_i x^i\right)\bar\pi_{\alpha_2}\right]\right\rbrace  \nonumber
\\&+&\frac{1}{\kappa}\Re\left\lbrace \int_\Sigma d^2\xi\left[\pi_{\alpha_1} f_{11}^{\alpha_1\beta_1}\bar\pi_{\beta_1}+\pi_{\alpha_1} f_{14}^{\alpha_1\beta_1}\bar\pi_{\beta_1}^*-\pi_{\alpha_2} f_{22}^{\alpha_2\beta_2}\bar\pi_{\beta_2}-\pi_{\alpha_2} f_{23}^{\alpha_2\beta_2}\bar\pi_{\beta_2}^*\right]\right\rbrace  \\ &+&\frac{1}{\kappa}\Re\left\lbrace \int_\Sigma d^2\xi\left[\pi_{\alpha_2} f_{21}^{\alpha_2\beta_1}\bar\pi_{\beta_1}-\pi_{\alpha_1} f_{12}^{\alpha_1\beta_2}\bar\pi_{\beta_2}+\pi_{\alpha_2} f_{24}^{\alpha_2\beta_1}\bar\pi_{\beta_1}^*-\pi_{\alpha_1} f_{13}^{\alpha_1\beta_2}\bar\pi_{\beta_2}^*\right]\right\rbrace\,
,\nonumber
\end{eqnarray}
where $\Re$ means real part of some complex number and
${}^*$ means complex conjugation.

In Table 1 we
summarize the list of the background fields of the type IIB superstring
theory in $D=10$ dimensional space-time, fields living on the
$D5$-brane and the rest fields.

\begin{table}[h]
\begin{tabular}{|c|c|c|c|}\hline
Sector  & $x^\mu\,(\mu=0,1,\dots,9)$    & $x^i\,(i=0,1,\dots,5)$ & $x^a\, (a=6,7,8,9)$ \\ \hline\hline
\raisebox{-1.5ex}[0pt]{NS-NS} & $G_{\mu\nu}$ & $G_{ij}$ & $G_{ia}=0\, ,G_{ab}$(decoupled) \\
      & $B_{\mu\nu}$ & $B_{ij}$ & $B_{ia}=B_{ab}=0$\\
      & $\Phi=0$ &  & \\ \hline
\raisebox{-1.5ex}[0pt]{NS-R} & $\Psi^\alpha_\mu$ & $\Psi^{\alpha_1}_i\, ,\Psi^{\alpha_2}_i$ & $\Psi^{\alpha}_a=0$ \\
                             & $\bar\Psi^\alpha_\mu$ & $\bar\Psi^{\alpha_1}_i\, ,\bar\Psi^{\alpha_2}_i$ & $\bar\Psi^{\alpha}_a=0$ \\ \hline
\raisebox{-1.5ex}[0pt]{R-R} & \raisebox{-1.5ex}[0pt]{$F^{\alpha\beta}$} & $f_{11}^{\alpha_1\beta_1}, f_{22}^{\alpha_2\beta_2}, f_{14}^{\alpha_1\beta_1}, f_{23}^{\alpha_2\beta_2}$ & \\
                            &  & $f_{12}^{\alpha_1\beta_2}, f_{21}^{\alpha_2\beta_1}, f_{13}^{\alpha_1\beta_2}, f_{24}^{\alpha_2\beta_1}$ & \\ \hline
\end{tabular}
\caption{Background fields of the type IIB theory: the complete set, part living on $D5$-brane and the rest fields eliminated from the theory.}
\end{table}

\section{Canonical analysis of type IIB superstring theory with $D5$-brane}
\setcounter{equation}{0}

In this section we will perform canonical analysis of pure spinor
formulation of type IIB theory with embedded $D5$-brane.

\subsection{Canonical Hamiltonian}

Momentum canonically conjugated to $x^i$  has the form
\begin{equation}
\pi_i\equiv \frac{\partial \mathcal L}{\partial \dot
x^i}=\kappa(G_{ij}\dot x^j-2B_{ij}x'^j)+2\Re \left(-\pi_{\alpha_1}
\Psi^{\alpha_1}_i-\pi_{\alpha_2}
\Psi^{\alpha_2}_i+\bar\Psi^{\alpha_1}_{i} \bar\pi_{\alpha_1}+\bar\Psi^{\alpha_2}_{i} \bar\pi_{\alpha_2}\right)\, ,
\end{equation}
while by definition the momenta $\pi_{\alpha_1}$, $\pi_{\alpha_2}$,
$\bar\pi_{\alpha_1}$ and $\bar\pi_{\alpha_2}$ are canonically
conjugated to the coordinates $\theta^{\alpha_1}$, $\theta^{\alpha_2}$,
$\bar\theta^{\alpha_1}$ and $\bar\theta^{\alpha_2}$, respectively.
In Appendix C we discussed this connection between
complex coordinates and momenta.

According to the definition of canonical Hamiltonian
\begin{equation}
\mathcal
H_c=\dot x^i \pi_i+2\Re\left(  \dot \theta^{\alpha_1} \pi_{\alpha_1}+\dot
{\bar\theta}^{\alpha_1}\bar\pi_{\alpha_1}+\dot \theta^{\alpha_2} \pi_{\alpha_2}+\dot
{\bar\theta}^{\alpha_2}\bar\pi_{\alpha_2}\right)  -\mathcal L\, ,\nonumber
\end{equation}
we have
\begin{equation}\label{eq:initialham}
H_c=\int d\sigma \mathcal H_c\, ,\quad \mathcal H_c=T_--T_+\, ,\quad T_{\pm} =t_{\pm}-\tau_{\pm}\, ,
\end{equation}
where
\begin{eqnarray}
t_{\pm}&=&\mp\frac{1}{4\kappa}G^{ij}I_{\pm i}I_{\pm j}\, ,\nonumber \\ I_{\pm i}&=&\pi_i+2\kappa \Pi_{\pm
ij}x'^j+2\Re\left(\pi_{\alpha_1} \Psi^{\alpha_1}_i+\pi_{\alpha_2} \Psi^{\alpha_2}_i-\bar\Psi^{\alpha_1}_i
\bar\pi_{\alpha_1}-\bar\Psi^{\alpha_2}_i
\bar\pi_{\alpha_2}\right),\nonumber \\
\tau_+&=&2\Re\left[ \left(\theta'^{\alpha_1}+\Psi^{\alpha_1}_i x'^i\right)
\pi_{\alpha_1}+\left(\theta'^{\alpha_2}+\Psi^{\alpha_2}_i x'^i\right)\pi_{\alpha_2}\right]\nonumber \\&-&\frac{1}{2\kappa}\Re\left( \pi_{\alpha_1} f_{11}^{\alpha_1\beta_1}\bar\pi_{\beta_1}+\pi_{\alpha_1} f_{14}^{\alpha_1\beta_1}\bar\pi_{\beta_1}^*-\pi_{\alpha_2} f_{22}^{\alpha_2\beta_2}\bar\pi_{\beta_2}-\pi_{\alpha_2} f_{23}^{\alpha_2\beta_2}\bar\pi_{\beta_2}^*\right)\nonumber \\ &-&\frac{1}{2\kappa}\Re\left( \pi_{\alpha_2} f_{21}^{\alpha_2\beta_1}\bar\pi_{\beta_1}-\pi_{\alpha_1} f_{12}^{\alpha_1\beta_2}\bar\pi_{\beta_2}+\pi_{\alpha_2} f_{24}^{\alpha_2\beta_1}\bar\pi_{\beta_1}^*-\pi_{\alpha_1} f_{13}^{\alpha_1\beta_2}\bar\pi_{\beta_2}^*\right)\, ,\nonumber \\
\tau_-&=&2\Re\left[ \left(
\bar\theta'^{\alpha_1}+\bar\Psi^{\alpha_1}_i x'^i\right)
\bar\pi_{\alpha_1}+\left(\bar\theta'^{\alpha_2}+\bar\Psi^{\alpha_2}_i x'^i\right)\bar\pi_{\alpha_2}\right]\nonumber
\\&+&\frac{1}{2\kappa}\Re\left( \pi_{\alpha_1} f_{11}^{\alpha_1\beta_1}\bar\pi_{\beta_1}+\pi_{\alpha_1} f_{14}^{\alpha_1\beta_1}\bar\pi_{\beta_1}^*-\pi_{\alpha_2} f_{22}^{\alpha_2\beta_2}\bar\pi_{\beta_2}-\pi_{\alpha_2} f_{23}^{\alpha_2\beta_2}\bar\pi_{\beta_2}^*\right)\nonumber \\ &+&\frac{1}{2\kappa}\Re\left( \pi_{\alpha_2} f_{21}^{\alpha_2\beta_1}\bar\pi_{\beta_1}-\pi_{\alpha_1} f_{12}^{\alpha_1\beta_2}\bar\pi_{\beta_2}+\pi_{\alpha_2} f_{24}^{\alpha_2\beta_1}\bar\pi_{\beta_1}^*-\pi_{\alpha_1} f_{13}^{\alpha_1\beta_2}\bar\pi_{\beta_2}^*\right) \, .
\label{eq:struja}
\end{eqnarray}

Using the standard Poisson bracket algebra
\begin{equation}\label{eq:xp}
\{x^{i}(\sigma)\, ,\pi_j(\bar\sigma)\}=\delta^{i}{}_j
\delta(\sigma-\bar\sigma)\, ,
\end{equation}
\begin{eqnarray}\label{eq:spaf}
\{\theta^{\alpha_1}(\sigma)\,
,\pi_{\beta_{1}}(\bar\sigma)\}&=&-\delta^{\alpha_1}{}_{\beta_{1}}\delta(\sigma-\bar\sigma)\,
,\quad \{\bar\theta^{\alpha_1}(\sigma)\,
,\bar\pi_{\beta_{1}}(\bar\sigma)\}=-\delta^{\alpha_1}{}_{\beta_{1}}\delta(\sigma-\bar\sigma)\,
,\nonumber \\
\{\theta^{\alpha_2}(\sigma)\,
,\pi_{\beta_{2}}(\bar\sigma)\}&=&-\delta^{\alpha_2}{}_{\beta_{2}}\delta(\sigma-\bar\sigma)\,
,\quad \{\bar\theta^{\alpha_2}(\sigma)\,
,\bar\pi_{\beta_{2}}(\bar\sigma)\}=-\delta^{\alpha_2}{}_{\beta_{2}}\delta(\sigma-\bar\sigma)\,
,
\end{eqnarray}
we calculate the algebra of currents
\begin{equation}
\{I_{\pm i}(\sigma)\, ,I_{\pm j}(\bar\sigma)\}=\pm 2\kappa
G_{ij}\delta'\, ,\quad \{I_{\pm i}(\sigma)\, ,I_{\mp
j}(\bar\sigma)\}=0\, .\quad \left[\delta'\equiv\partial_\sigma
\delta(\sigma-\bar\sigma)\right]
\end{equation}

With the help of this algebra we find that components $T_{\pm}$ satisfy
Virasoro algebra
\begin{equation}
\left\lbrace T_{\pm}(\sigma)\, ,T_{\pm}(\bar\sigma)\right\rbrace=
-\left[ T_{\pm}(\sigma)+T_{\pm}(\bar\sigma)\right] \delta'\,
,\quad  \left\lbrace T_{\pm}(\sigma)\,
,T_{\mp}(\bar\sigma)\right\rbrace=0\, .
\end{equation}
The Poisson bracket between canonical Hamiltonian and current
$I_{\pm i}$ is proportional to its sigma derivative
\begin{equation}\label{eq:pzhj}
\{H_c\, ,I_{\pm i}\}=\mp I'_{\pm i}\, .
\end{equation}

\subsection{Boundary conditions as canonical constraints}

Following method of Ref.\cite{radepjc}, using canonical approach,
we will derive boundary conditions directly in terms of canonical
variables. Varying Hamiltonian $H_c$ we obtain
\begin{equation}
\delta H_c=\delta H_c^{(R)}-\left[ \gamma_i^{(0)}\delta
x^i+2\Re\left( \pi_{\alpha_1}\delta\theta^{\alpha_1}+\delta
\bar\theta^{\alpha_1}\bar\pi_{\alpha_1}\right)+2\Re\left(
\pi_{\alpha_2}\delta\theta^{\alpha_2}+\delta
\bar\theta^{\alpha_2}\bar\pi_{\alpha_2}\right) \right]
\Big|_0^\pi\, ,
\end{equation}
where $\delta H_c^{(R)}$ is regular term of the form
\begin{eqnarray}
\delta H_c^{(R)}&=&\int d\sigma \left[ A_i \delta x^i+B^i \delta
\pi_i+2\Re\left( C_{\alpha_1}
\delta\theta^{\alpha_1}+D_{\alpha_1}\delta
\bar\theta^{\alpha_1}+E^{\alpha_1}\delta\pi_{\alpha_1}+F^{\alpha_1}\delta
\bar\pi_{\alpha_1}\right)\right]\nonumber \\ &+&2\Re\int d\sigma
\left( C_{\alpha_2} \delta\theta^{\alpha_2}+D_{\alpha_2}\delta
\bar\theta^{\alpha_2}+E^{\alpha_2}\delta\pi_{\alpha_2}+F^{\alpha_2}\delta
\bar\pi_{\alpha_2}\right) \, ,
\end{eqnarray}
and
\begin{equation}
\gamma_{i}^{(0)}=\Pi_{+ i}{}^j I_{- j}+\Pi_{- i}{}^j I_{+
j}+2\Re\left( \pi_{\alpha_1}
\Psi^{\alpha_1}_i+\pi_{\alpha_2}
\Psi^{\alpha_2}_i+\bar\Psi^{\alpha_1}_i \bar\pi_{\alpha_1}+\bar\Psi^{\alpha_2}_i \bar\pi_{\alpha_2}\right)
\, .
\end{equation}
As a time translation generator Hamiltonian must have
well defined derivatives with respect to its variables. Consequently, boundary
term has to vanish and we obtain
\begin{equation}\label{eq:BC}
\left[  \gamma_i^{(0)}\delta x^i+2\Re\left(
\pi_{\alpha_1}\delta\theta^{\alpha_1}+\delta
\bar\theta^{\alpha_1}\bar\pi_{\alpha_1}\right)+2\Re\left(
\pi_{\alpha_2}\delta\theta^{\alpha_2}+\delta
\bar\theta^{\alpha_2}\bar\pi_{\alpha_2}\right)\right] \Big
|_0^\pi=0\, .
\end{equation}

For bosonic coordinates $x^i$ we choose Neumann boundary conditions, implying
\begin{equation}\label{eq:gamami}
\gamma_i^{(0)}\big |_0^\pi=0\, ,
\end{equation}
while for fermionic coordinates we choose
\begin{equation}\label{eq:bcf1}
(\theta^{\alpha_1}-\bar\theta^{\alpha_1})\Big |_0^\pi=0\, ,\quad (\theta^{\alpha_2}+\bar\theta^{\alpha_2})\Big |_0^\pi=0\, ,
\end{equation}
which produces additional boundary conditions
\begin{equation}\label{eq:bcf2}
(\pi_{\alpha_1}-\bar\pi_{\alpha_1})\big |_0^\pi=0\, , \quad (\pi_{\alpha_2}+\bar\pi_{\alpha_2})\big |_0^\pi=0\, .
\end{equation}
According with Refs.\cite{radepjc,BNBS,BNBS2}, we will treat the expressions
(\ref{eq:gamami})-(\ref{eq:bcf2}) as canonical constraints.

\subsection{Consistency of bosonic constraints}

Using Eq.(\ref{eq:pzhj}) and standard Poisson algebra, the
consistency procedure for $\gamma_i^{(0)}$ produces an infinite
set of constraints $\gamma_{i}^{(n)}\,(n=1,2,\dots)$ with
\begin{eqnarray}
&{}&\gamma_{i}^{(n)}\equiv\{H_c\, ,\gamma_i^{(n-1)}\}=\Pi_{+
i}{}^j \partial_\sigma^{(n)}I_{- j}+(-1)^n\Pi_{- i}{}^j
\partial_\sigma^{(n)}I_{+ j}\nonumber \\ &+&  2\Re \left[(-1)^n
\partial_\sigma^{(n)}\pi_{\alpha_1}
\Psi^{\alpha_1}_i+(-1)^n
\partial_\sigma^{(n)}\pi_{\alpha_2}
\Psi^{\alpha_2}_i+\bar\Psi^{\alpha_1}_i
\partial_\sigma^{(n)}\bar\pi_{\alpha_1}+\bar\Psi^{\alpha_2}_i
\partial_\sigma^{(n)}\bar\pi_{\alpha_2}\right]  \, .
\end{eqnarray}
They can be rewritten in the compact $\sigma$-dependent form
\begin{eqnarray}
\Gamma_i(\sigma)&\equiv&\sum_{n=0}^\infty
\frac{\sigma^n}{n!}\gamma_i^{(n)}(\sigma=0)=\Pi_{+ i}{}^j I_{-
j}(\sigma)+\Pi_{- i}{}^j I_{+ j}(-\sigma)\nonumber \\
&+&2\Re \left[\pi_{\alpha_1}(-\sigma)\Psi^{\alpha_1}_{i}+\pi_{\alpha_2}(-\sigma)\Psi^{\alpha_2}_{i}+\bar\Psi^{\alpha_1}_i
\bar\pi_{\alpha_1}(\sigma)+\bar\Psi^{\alpha_2}_i
\bar\pi_{\alpha_2}(\sigma)\right]\nonumber \\ &=&\tilde
p_i+2(BG^{-1})_i{}^j p_j-\kappa G^{eff}_{ij}\tilde q'^j+ 4\Re\left[\Pi_{+ i}{}^j P_s \pi_{\alpha_1}
(\sigma)\Psi^{\alpha_1}_j+\Pi_{+ i}{}^j P_s \pi_{\alpha_2}
(\sigma)\Psi^{\alpha_2}_j\right]\nonumber \\ &-&4\Re\left[\Pi_{- i}{}^j \bar\Psi^{\alpha_1}_j P_s
\bar\pi_{\alpha_1}(\sigma)+\Pi_{- i}{}^j \bar\Psi^{\alpha_2}_j P_s
\bar\pi_{\alpha_2}(\sigma)\right]\, .
\end{eqnarray}
Here we introduced new variables, even and odd
under world-sheet parity transformation $\Omega:\sigma\to -\sigma$. For
bosonic variables we use standard notation \cite{BNBS,BNBS2}
\begin{equation}\label{eq:bv1}
q^i(\sigma)= P_s x^i(\sigma)\equiv\frac{1}{2}\left [
x^i(\sigma)+x^i(-\sigma)\right ]\, ,\quad \tilde
q^i(\sigma)= P_a x^i(\sigma)\equiv\frac{1}{2}\left [
x^i(\sigma)-x^i(-\sigma)\right ]\, ,\nonumber
\end{equation}
\begin{equation}\label{eq:bv2}
p_i(\sigma)= P_s \pi_i(\sigma)\equiv\frac{1}{2}\left [
\pi_i(\sigma)+\pi_i(-\sigma)\right ]\, ,\quad \tilde
p_i(\sigma)= P_a \pi_i(\sigma)\equiv\frac{1}{2}\left [
\pi_i(\sigma)-\pi_i(-\sigma)\right ]\, ,
\end{equation}
while for fermionic ones we explicitly use
the projectors on $\Omega$ even and odd parts
\begin{equation}\label{eq:PsPa}
P_s=\frac{1}{2}(1+\Omega)\, ,\quad P_a=\frac{1}{2}(1-\Omega)\, .
\end{equation}

\subsection{Consistency of fermionic constraints}

The complete consistency procedure is given in Appendix D. Here we write only the form of the fermionic constraints after Dirac consistency procedure.

We start consistency procedure for fermionic constraints, (\ref{eq:bcf1})
and (\ref{eq:bcf2}), applying this procedure to
the variables
$$A^{(0)}=(\theta^{\alpha_1}, \theta^{\alpha_2}, \bar\theta^{\alpha_1}, \bar\theta^{\alpha_2}, \pi_{\alpha_1}, \pi_{\alpha_2}, \bar\pi_{\alpha_1}, \bar\pi_{\alpha_2})\, ,$$
and obtain an infinite set of the constraints $A^{(n)}\;(n=0,1,2,3,\dots)$. The complete set of constraints following from boundary conditions (\ref{eq:bcf1}) and Dirac canonical procedure, in compact notation at $\sigma=0$ has the form
\begin{eqnarray}
\Gamma^{\alpha_1}(\sigma)=
\Theta^{\alpha_1}(\sigma)-\bar\Theta^{\alpha_1}(\sigma)\, ,\quad \Gamma^{\alpha_2}(\sigma)=
\Theta^{\alpha_2}(\sigma)+\bar\Theta^{\alpha_2}(\sigma)\, ,
\end{eqnarray}
where right-hand side variables are defined in (\ref{eq:Fialfa})-(\ref{eq:barFialfa}).
Similarly, from (\ref{eq:bcf2}), we have
\begin{eqnarray}
\Gamma_{\alpha_1}^\pi(\sigma)&\equiv&
\Pi_{\alpha_1}(\sigma)-\bar\Pi_{\alpha_1}(\sigma)=\pi_{\alpha_1}(-\sigma)-\bar\pi_{\alpha_1}(\sigma)\,
,\nonumber \\ \Gamma_{\alpha_2}^\pi(\sigma)&\equiv&
\Pi_{\alpha_2}(\sigma)+\bar\Pi_{\alpha_2}(\sigma)=\pi_{\alpha_2}(-\sigma)+\bar\pi_{\alpha_2}(\sigma)\, ,
\end{eqnarray}
where $\Pi_{\alpha_1}$, $\Pi_{\alpha_2}$, $\bar\Pi_{\alpha_1}$ and $\bar\Pi_{\alpha_2}$ are defined in Eq.(\ref{eq:Pialfa}).

For all bosonic and fermionic constraints we also apply the
consistency procedure at $\sigma =\pi$ and obtain similar
expressions, where all variables depending on $-\sigma$ are
replaced by the same variables depending on $2\pi-\sigma$. That
set of constraints is solved by $2\pi$ periodicity of all
canonical variables as well as in Refs.\cite{BNBS,BNBS2}.

\subsection{Classification of constraints}

Let us denote all constraints with $\Gamma_A=(\Gamma_i\,
,\Gamma^{\alpha_1}\, ,\Gamma^{\alpha_2}\, ,\Gamma_{\alpha_1}^\pi\, ,\Gamma_{\alpha_2}^\pi)$. From
\begin{equation}
\left\lbrace H_c\, ,\Gamma_A\right\rbrace=\Gamma_A'\approx0\, ,
\end{equation}
it follows that all constraints weakly commute with canonical
Hamiltonian, so there are no more constraints in the theory and
the consistency procedure is completed.

We are going to classify the constraints. For practical reasons,
in order to classify the constraints easier, let us first consider the
quantities $\Gamma'^{\alpha_1}$ and $\Gamma'^{\alpha_2}$ instead
$\Gamma^{\alpha_1}$ and $\Gamma^{\alpha_2}$, as constraints in the
theory. The algebra of the constraints
${}^\star{}\Gamma_A=(\Gamma_i\, ,\Gamma'^{\alpha_1}\,
,\Gamma'^{\alpha_2}\, ,\Gamma_{\alpha_1}^\pi\,
,\Gamma_{\alpha_2}^\pi)$ has the form
\begin{equation}\label{eq:algebraveza}
\left\lbrace {}^\star\Gamma_A\,
,{}^\star\Gamma_B\right\rbrace=M_{AB}\delta'\, ,
\end{equation}
where the supermatrix $M_{AB}$ is given by the expression
\begin{eqnarray}\label{eq:algebrav}
M_{AB}&=&\left(
\begin{array}{cc}
M_1 & M_2\\
M_3 & M_4
\end{array}\right)=\left(
\begin{array}{c|ccccc}
(M_1)_{ij} & M_i{}^{\gamma_1} & M_i{}^{\gamma_2} & \overline M_{i\delta_1} & \overline M_{i\delta_2}\\ \hline
M^{\alpha_1}{}_j & M^{\alpha_1\gamma_{1}} & M^{\alpha_1\gamma_{2}} & \overline M^{\alpha_1}{}_{\delta_{1}} & \overline M^{\alpha_1}{}_{\delta_{2}} \\
M^{\alpha_2}{}_j & M^{\alpha_2\gamma_{1}} & M^{\alpha_2\gamma_{2}} & \overline M^{\alpha_2}{}_{\delta_{1}} & \overline M^{\alpha_2}{}_{\delta_{2}} \\
\overline M_{\beta_1 j} & \overline M_{\beta_1}{}^{\gamma_{1}} & \overline M_{\beta_1}{}^{\gamma_{2}} & \tilde
M_{\beta_1\delta_{1}} & \tilde
M_{\beta_1\delta_{2}}\\
\overline M_{\beta_2 j} & \overline M_{\beta_2}{}^{\gamma_{1}} & \overline M_{\beta_2}{}^{\gamma_{2}} & \tilde
M_{\beta_2\delta_{1}} & \tilde
M_{\beta_2\delta_{2}}
\end{array}\right)\nonumber \\  \\&=&\left(
\begin{array}{c|ccccc}
-\kappa G_{ij}^{eff} & -2 (\Psi^{eff})_i{}^{\gamma_1} & -2(\Psi^{eff})_i{}^{\gamma_2} & 0 & 0\\ \hline
-2(\Psi^{eff})^{\alpha_1}{}_j & \frac{1}{\kappa}(f_{11}^{eff})^{\alpha_1\gamma_{1}} & \frac{1}{\kappa}(f_{12}^{eff})^{\alpha_1\gamma_{2}} & -2\delta^{\alpha_1}{}_{\delta_{1}} & 0 \\
-2(\Psi^{eff})^{\alpha_2}{}_j & -\frac{1}{\kappa}(f_{12}^{eff})^{\alpha_2\gamma_{1}} & \frac{1}{\kappa}(f_{22}^{eff})^{\alpha_2\gamma_{2}} & 0 & -2\delta^{\alpha_2}{}_{\delta_{2}} \\
0 & -2\delta_{\beta_1}{}^{\gamma_{1}} & 0 & 0 & 0\\
0 & 0 & -2\delta_{\beta_2}{}^{\gamma_{2}} & 0 & 0
\end{array}\right)\, .\nonumber
\end{eqnarray}
To the fields appearing in matrix $M_{AB}$ we will refer as the effective background fields. They are defined as
\begin{eqnarray}\label{eq:effbackground}
G^{eff}_{ij}&=&G_{ij}-4B_{i k}G^{kl}B_{l j}\,
,\nonumber
\\(\Psi_{eff})^{\alpha_1}_{i}&=&\frac{1}{2}\Psi^{\alpha_1}_{+ i}+B_{ik}G^{kj}
\Psi^{\alpha_1}_{- j}\, ,\quad
(\Psi_{eff})^{\alpha_2}_{i}=\frac{1}{2}\Psi^{\alpha_2}_{-
i}+B_{ik}G^{kj} \Psi^{\alpha_2}_{+ j}\, ,\nonumber \\
(f^{eff}_{11})^{\alpha_1\beta_1}&=&(f_{11}^a)^{\alpha_1\beta_1}-\Psi^{\alpha_1}_{-i}G^{ij}\Psi^{\beta_1}_{-j}\,
,\quad
(f^{eff}_{22})^{\alpha_2\beta_2}=(f^a_{22})^{\alpha_2\beta_2}-\Psi^{\alpha_2}_{+i}G^{ij}\Psi^{\beta_2}_{+j}\,
,\nonumber \\
(f_{12}^{eff})^{\alpha_1\beta_2}&=&\frac{1}{2}\left(f_{12}^{\alpha_1\beta_2}-f_{21}^{\beta_2\alpha_1} \right) - \Psi^{\alpha_1}_{- i} G^{ij}\Psi^{\beta_2}_{+j}\, ,
\end{eqnarray}
where
\begin{eqnarray}\label{eq:saF}
\Psi^{\alpha_1}_{\pm
i}&=&\Psi^{\alpha_1}_i \pm \bar\Psi^{\alpha_1}_i\, ,\nonumber \\
(f^s)^{\alpha_1\beta_1}&=&\frac{1}{2}(f^{\alpha_1\beta_1}+f^{\beta_1\alpha_1})\,
,\quad
(f^a)^{\alpha_1\beta_1}=\frac{1}{2}(f^{\alpha_1\beta_1}-f^{\beta_1\alpha_1})\,
.
\end{eqnarray}
The matrix $M_{AB}$ is independent of the field strengths
$f_{14}^{\alpha_1\beta_2}$, $f_{23}^{\alpha_2\beta_2}$,
$f_{13}^{\alpha_1\beta_2}$ and $f_{24}^{\alpha_2\beta_1}$ because
boundary conditions depend on complex conjugated momenta $\pi^*$
but not on the corresponding coordinates $\theta^*$.

From the definition of superdeterminant
\begin{equation}
s\det M_{AB}=\frac{\det(M_1-M_2 M_4^{-1}M_3)}{\det M_4}\, ,
\end{equation}
and using the fact that
\begin{equation}
(M_2 M_4^{-1}M_3)_{ij}=0\, ,\quad \det M_4=const.\, ,
\end{equation}
we obtain
\begin{equation}\label{eq:detM}
s\det M_{AB}\sim\det G^{eff}_{ij}\, .
\end{equation}

Because we assume that effective metric $G_{ij}^{eff}$ is
nonsingular, we conclude that constraints ${}^\star\Gamma_A$ are
of the second class.

The original constraints $\Gamma^{\alpha_1}$ and
$\Gamma^{\alpha_2}$ have the form (see Appendix D)
$$
\Gamma^{\alpha_1}=(\theta^{\alpha_1}-\bar\theta^{\alpha_1})|_{\sigma=0}+\sum_{n=1}^{\infty}\frac{\sigma^n}{n!}(\Theta^{\alpha_1}_{(n)}-\bar\Theta^{\alpha_1}_{(n)})(\sigma=0)\,
,
$$
$$
\Gamma^{\alpha_2}=(\theta^{\alpha_2}+\bar\theta^{\alpha_2})|_{\sigma=0}+\sum_{n=1}^{\infty}\frac{\sigma^n}{n!}(\Theta^{\alpha_1}_{(n)}+\bar\Theta^{\alpha_2}_{(n)})(\sigma=0)\,
.
$$
The second class constraints $\Gamma'^{\alpha_1}$ and
$\Gamma'^{\alpha_2}$ are effectively the derivatives of the parts
under the sums. It is easy to check that zero modes,
$(\theta^{\alpha_1}-\bar\theta^{\alpha_1})|_{\sigma=0}$ and
$(\theta^{\alpha_2}+\bar\theta^{\alpha_2})|_{\sigma=0}$, are also
second class constraints, and consequently, the complete set
of the constraints, $\Gamma_A$, is of the second class.

The reason for having two steps in this consideration is that
Poisson brackets of constraints ${}^\star\Gamma_A$ closed on
$\delta'$ function while those with zero modes closed on $\delta$
function.

\section{Solution of the constraints}
\setcounter{equation}{0}

The solution of the second class constraints originating from boundary conditions,
$\Gamma_i=0$, $\Gamma^{\alpha_1}=0$, $\Gamma^{\alpha_2}=0$, $\Gamma_{\alpha_1}^\pi=0$  and $\Gamma_{\alpha_2}^\pi=0$, has
the form
\begin{equation}\label{eq:resenjex}
x^{i}(\sigma)=q^i-2\Theta^{ij}\int_0^\sigma d\sigma_1
p_j+4\Re\left(\Theta^{i\alpha_1}\int_0^\sigma d\sigma_1 p_{\alpha_1}+\Theta^{i\alpha_2}\int_0^\sigma d\sigma_1 p_{\alpha_2}\right)\,
,\quad \pi_i=p_i\, ,
\end{equation}
\begin{eqnarray}\label{eq:resenje1}
\theta^{\alpha_1}(\sigma)&=&\frac{1}{2}(\xi^{\alpha_1}+\tilde\xi^{\alpha_1})-\Theta^{i\alpha_1}\int_0^\sigma
d\sigma_1 p_i -\Theta^{\alpha_1\beta_{1}}\int_0^\sigma d\sigma_1
p_{\beta_{1}}-\Theta^{\alpha_1\beta_{2}}\int_0^\sigma d\sigma_1
p_{\beta_{2}}\nonumber \\&-&{}^\star\Theta^{\alpha_1\beta_1}\int_0^\sigma d\sigma_1
p^*_{\beta_{1}}-{}^\star\Theta^{\alpha_1\beta_2}\int_0^\sigma d\sigma_1
p^*_{\beta_{2}}\, ,\nonumber \\ \pi_{\alpha_1}&=&p_{\alpha_1}+\tilde p_{\alpha_1}\, ,
\end{eqnarray}
\begin{eqnarray}\label{eq:resenje12}
\theta^{\alpha_2}(\sigma)&=&\frac{1}{2}(\xi^{\alpha_2}+\tilde\xi^{\alpha_2})-\Theta^{i\alpha_2}\int_0^\sigma
d\sigma_1 p_i-\Theta^{\alpha_2\beta_1}\int_0^\sigma d\sigma_1
p_{\beta_{1}}-\Theta^{\alpha_2\beta_2}\int_0^\sigma d\sigma_1
p_{\beta_{2}}\nonumber \\ &-&{}^\star
\Theta^{\alpha_2\beta_1}\int_0^\sigma d\sigma_1
p^*_{\beta_{1}}-{}^\star \Theta^{\alpha_2\beta_2}\int_0^\sigma
d\sigma_1 p^*_{\beta_{2}}\, ,\nonumber \\
\pi_{\alpha_2}&=&p_{\alpha_2}+\tilde p_{\alpha_2}\, ,
\end{eqnarray}
\begin{eqnarray}\label{eq:resenje2}
&{}&\bar\theta^{\alpha_1}(\sigma)=\theta^{\alpha_1}(\sigma)-\tilde \xi^{\alpha_1}(\sigma)=\frac{1}{2}(\xi^{\alpha_1}-\tilde\xi^{\alpha_1})-\Theta^{i\alpha_1}\int_0^\sigma
d\sigma_1 p_i \nonumber \\ &-& \Theta^{\alpha_1\beta_{1}}\int_0^\sigma d\sigma_1
p_{\beta_{1}}- \Theta^{\alpha_1\beta_{2}}\int_0^\sigma d\sigma_1
p_{\beta_{2}}-{}^\star\Theta^{\alpha_1\beta_1}\int_0^\sigma d\sigma_1
p^*_{\beta_{1}}-{}^\star\Theta^{\alpha_1\beta_2}\int_0^\sigma d\sigma_1
p^*_{\beta_{2}}\, ,\nonumber \\ &{}&\bar\pi_{\alpha_1}=p_{\alpha_1}-\tilde p_{\alpha_1}\, ,
\end{eqnarray}
\begin{eqnarray}\label{eq:resenje22}
&{}&\bar\theta^{\alpha_2}(\sigma)=-\theta^{\alpha_2}(\sigma)+\tilde\xi^{\alpha_2}(\sigma)=-\frac{1}{2}(\xi^{\alpha_2}-\tilde\xi^{\alpha_2})+\Theta^{i\alpha_2}\int_0^\sigma
d\sigma_1 p_i\nonumber \\&+&\Theta^{\alpha_2\beta_1}\int_0^\sigma d\sigma_1
p_{\beta_{1}}+\Theta^{\alpha_2\beta_2}\int_0^\sigma d\sigma_1
p_{\beta_{2}}+{}^\star
\Theta^{\alpha_2\beta_1}\int_0^\sigma d\sigma_1
p^*_{\beta_{1}}+{}^\star \Theta^{\alpha_2\beta_2}\int_0^\sigma
d\sigma_1 p^*_{\beta_{2}}\, ,\nonumber \\
&{}&\bar\pi_{\alpha_2}=-p_{\alpha_2}+\tilde p_{\alpha_2}\, ,
\end{eqnarray}
in terms of the bosonic $\Omega$ even and odd variables given in (\ref{eq:bv2}) and the fermionic variables
\begin{eqnarray}
\frac{1}{2}\xi^{\alpha_1}&\equiv&
P_s \theta^{\alpha_1}=P_s \bar\theta^{\alpha_1}\, ,\quad
\tilde\xi^{\alpha_1}\equiv P_a(\theta^{\alpha_1}-\bar\theta^{\alpha_1})\,
,\nonumber \\ \frac{1}{2}\xi^{\alpha_2}&\equiv&
P_s \theta^{\alpha_2}=-P_s \bar\theta^{\alpha_2}\, ,\quad
\tilde\xi^{\alpha_2}\equiv P_a(\theta^{\alpha_2}+\bar\theta^{\alpha_2})\,
,\nonumber \\ p_{\alpha_1}&\equiv& P_s \pi_{\alpha_1}=P_s \bar\pi_{\alpha_1}\, ,\quad
\tilde p_{\alpha_1}\equiv P_a\pi_{\alpha_1}=-P_a \bar\pi_{\alpha_1}   \, ,\nonumber \\ p_{\alpha_2}&\equiv& P_s \pi_{\alpha_2}=-P_s \bar\pi_{\alpha_2}\, ,\quad
\tilde p_{\alpha_2}\equiv P_a\pi_{\alpha_2}=P_a \bar\pi_{\alpha_2}   \, .\label{eq:resenjepi}
\end{eqnarray}
The coefficients multiplying momenta are defined as
\begin{eqnarray}\label{eq:teta1}
\Theta^{ij}&=&-\frac{1}{\kappa}(G_{eff}^{-1}BG^{-1})^{ij}\,
, \\
\Theta^{i \alpha_1}&=&2\Theta^{ij}(\Psi_{eff})^{\alpha_1}_j-\frac{1}{2\kappa}G^{ij}\Psi^{\alpha_1}_{- j}\,
,\quad \Theta^{i \alpha_2}=2\Theta^{ij}(\Psi_{eff})^{\alpha_2}_j-\frac{1}{2\kappa}G^{ij}\Psi^{\alpha_2}_{+ j}\, ,
\end{eqnarray}
\begin{eqnarray}\label{eq:teta2}
\Theta^{\alpha_1\beta_{1}}&=&\frac{1}{2\kappa}(f^s_{11})^{\alpha_1 \beta_{1}}+4(\Psi_{eff})^{\alpha_1}_i
\Theta^{ij}(\Psi_{eff})^{\beta_{1}}_j-\frac{1}{\kappa}\Psi^{\alpha_1}_{- i}(G^{-1}BG^{-1})^{ij}\Psi^{\beta_{1}}_{- j}\nonumber \\ &+&\frac{G^{ij}}{\kappa}\left[ \Psi^{\alpha_1}_{- i}(\Psi_{eff})^{\beta_{1}}_j+\Psi^{\beta_{1}}_{- i}(\Psi_{eff})^{\alpha_{1}}_j
\right] \,
,
\end{eqnarray}
\begin{eqnarray}\label{eq:teta2222}
\Theta^{\alpha_1\beta_{2}}&=&\Theta^{\beta_2\alpha_1}=\frac{1}{4\kappa}(f_{12}^{\alpha_1\beta_2}+f_{21}^{\beta_2\alpha_1})-4(\Psi_{eff})^{\alpha_1}_i
\Theta^{ij}(\Psi_{eff})^{\beta_{2}}_j\nonumber \\ &-&\frac{1}{\kappa}\Psi^{\alpha_1}_{- i}(G^{-1}BG^{-1})^{ij}\Psi^{\beta_{2}}_{+ j}+\frac{G^{ij}}{\kappa}\left[ \Psi^{\alpha_1}_{- i}(\Psi_{eff})^{\beta_{2}}_j+\Psi^{\beta_{2}}_{+ i}(\Psi_{eff})^{\alpha_{1}}_j
\right] \,
,
\end{eqnarray}
\begin{eqnarray}\label{eq:teta21}
\Theta^{\alpha_2\beta_{2}}&=&\frac{1}{2\kappa}(f^s_{22})^{\alpha_2
\beta_{2}}+4(\Psi_{eff})^{\alpha_2}_i
\Theta^{ij}(\Psi_{eff})^{\beta_{2}}_j-\frac{1}{\kappa}\Psi^{\alpha_2}_{+
i}(G^{-1}BG^{-1})^{ij}\Psi^{\beta_{2}}_{+ j}\nonumber \\
&+&\frac{G^{ij}}{\kappa}\left[ \Psi^{\alpha_2}_{+
i}(\Psi_{eff})^{\beta_{2}}_j+\Psi^{\beta_{2}}_{+
i}(\Psi_{eff})^{\alpha_{2}}_j \right] \, ,
\end{eqnarray}
\begin{eqnarray}\label{eq:teta22}
{}^\star\Theta^{\alpha_1\beta_{1}}&=&\frac{1}{4\kappa}(f_{14}^{\alpha_1\beta_1}+f_{14}^{*\beta_1\alpha_1})+4(\Psi_{eff})^{\alpha_1}_i
\Theta^{ij}(\Psi_{eff})^{*\beta_{1}}_j-\frac{1}{\kappa}\Psi^{\alpha_1}_{- i}(G^{-1}BG^{-1})^{ij}\Psi^{*\beta_{1}}_{- j}\nonumber \\ &+&\frac{G^{ij}}{\kappa}\left[ \Psi^{\alpha_1}_{- i}(\Psi_{eff})^{*\beta_{1}}_j+\Psi^{*\beta_{1}}_{- i}(\Psi_{eff})^{\alpha_{1}}_j
\right] \,
,
\end{eqnarray}
\begin{eqnarray}
{}^\star\Theta^{\alpha_1\beta_{2}}&=&{}^\star \Theta^{\beta_2\alpha_1}=\frac{1}{4\kappa}(f_{13}^{\alpha_1\beta_2}+f_{24}^{*\beta_2\alpha_1})-4(\Psi_{eff})^{\alpha_1}_i
\Theta^{ij}(\Psi_{eff})^{*\beta_{2}}_j\nonumber \\&-&\frac{1}{\kappa}\Psi^{\alpha_1}_{- i}(G^{-1}BG^{-1})^{ij}\Psi^{*\beta_{2}}_{+ j}+\frac{G^{ij}}{\kappa}\left[ \Psi^{\alpha_1}_{- i}(\Psi_{eff})^{*\beta_{2}}_j+\Psi^{*\beta_{2}}_{+ i}(\Psi_{eff})^{\alpha_{1}}_j
\right] \,
,
\end{eqnarray}
\begin{eqnarray}
{}^\star\Theta^{\alpha_2\beta_{2}}&=&\frac{1}{4\kappa}(f_{23}^{\alpha_2\beta_2}+f_{23}^{*\beta_2\alpha_2})+4(\Psi_{eff})^{\alpha_2}_i
\Theta^{ij}(\Psi_{eff})^{*\beta_{2}}_j-\frac{1}{\kappa}\Psi^{\alpha_2}_{+
i}(G^{-1}BG^{-1})^{ij}\Psi^{*\beta_{2}}_{+ j}\nonumber \\
&+&\frac{G^{ij}}{\kappa}\left[ \Psi^{\alpha_2}_{+
i}(\Psi_{eff})^{*\beta_{2}}_j+\Psi^{*\beta_{2}}_{+
i}(\Psi_{eff})^{\alpha_{2}}_j \right] \, .
\end{eqnarray}

\section{Type I theory as effective theory of type IIB one}
\setcounter{equation}{0}

In this section we will find the effective theory, which is
defined as type IIB theory on the solution of boundary conditions.
Correlating it with type I theory we get relations between their
background fields.

\subsection{Effective Hamiltonian and Lagrangian of type IIB theory with $D5$-brane}

The initial theory is described by variables $x^i$,
$\theta^{\alpha_1}$, $\theta^{\alpha_2}$, $\bar\theta^{\alpha_1}$ and $\bar\theta^{\alpha_2}$. On the solution of
the constraints initial theory turns into effective one expressed
in terms of effective variables with well defined $\Omega$ parity
$q^i$, $\xi^{\alpha_1}$, $\xi^{\alpha_2}$, $\tilde \xi^{\alpha_1}$ and $\tilde \xi^{\alpha_2}$.

To obtain an effective Hamiltonian we have to put the solutions of
the constraints into the expression for canonical Hamiltonian.
First, we introduce effective current
\begin{eqnarray}\label{eq:efstruja}
\tilde I_{\pm i}&=& p_i \pm \kappa
G^{eff}_{ij}q'^j-4\Re\left[(\Psi_{eff})^{\alpha_1}_i
p_{\alpha_1}+(\Psi_{eff})^{\alpha_2}_i
p_{\alpha_2}\right]\nonumber \\ &+&4\Re\left[ \Pi_{\mp i}{}^j \Psi^{\alpha_1}_{- j}(p_{\alpha_1}\pm \tilde
p_{\alpha_1})+\Pi_{\mp i}{}^j \Psi^{\alpha_2}_{+ j}(p_{\alpha_2}\pm \tilde
p_{\alpha_2})\right]\, ,
\end{eqnarray}
and correlate it with the currents $I_{\pm i}$ (\ref{eq:struja})
\begin{equation}\label{eq:relacijaI}
I_{\pm i}=\pm 2 (\Pi_\pm G_{eff}^{-1})_i{}^j \tilde I_{\pm
j}\, .
\end{equation}
Now, substituting last relation and the solutions
(\ref{eq:resenjex})-(\ref{eq:resenje22}) into canonical Hamiltonian
(\ref{eq:initialham}), we get the expression for effective one
\begin{eqnarray}\label{eq:geffh}
{\mathcal H}^{eff}_c&=&\tilde t_{-}-\tilde
t_{+}\nonumber \\&+&2\Re\left[\Psi^{\alpha_1}_{- i}\left(
\frac{1}{\kappa}G_{eff}^{ij}\tilde
p_{\alpha_1}-2\Theta^{ij}p_{\alpha_1}\right) p_j+\Psi^{\alpha_2}_{+ i}\left(
\frac{1}{\kappa}G_{eff}^{ij}\tilde
p_{\alpha_2}-2\Theta^{ij}p_{\alpha_2}\right) p_j\right]\nonumber \\&+& 2\Re\left[ \left(\tilde
\xi'^{\alpha_1}+\Psi^{\alpha_1}_{- i} q'^i\right)p_{\alpha_1} + \left(
\xi'^{\alpha_1}+\Psi^{\alpha_1}_{+ i} q'^i\right )  \tilde p_{\alpha_1}\right]\nonumber \\&+&2\Re\left[\left( \tilde\xi'^{\alpha_2}+\Psi^{\alpha_2}_{+ i} q'^i\right)p_{\alpha_2}+\left(\xi'^{\alpha_2}+\Psi^{\alpha_2}_{- i} q'^i\right)\tilde p_{\alpha_2}\right]
\nonumber \\&-&\frac{1}{\kappa}\Re\left[ p_{\alpha_1}\left(
 (f^a_{11})^{\alpha_1\beta_{1}}+4\kappa
\Psi^{\alpha_1}_{- i}\Theta^{i\beta_{1}}\right)
p_{\beta_{1}}\right]+\frac{1}{\kappa}\Re\left( \tilde p_{\alpha_1} (f_{11}^a)^{\alpha_1\beta_1}\tilde p_{\beta_1}\right)\nonumber \\&-&\frac{1}{\kappa}\Re\left[ p_{\alpha_2}\left(
 (f^a_{22})^{\alpha_2\beta_{2}}+4\kappa
\Psi^{\alpha_2}_{+ i}\Theta^{i\beta_{2}}\right)p_{\beta_2}
\right]+\frac{1}{\kappa}\Re\left[ \tilde
p_{\alpha_2}(f^a_{22})^{\alpha_2\beta_2}\tilde p_{\beta_2}
\right]\nonumber
\\&+&\frac{4}{\kappa}\Re \left[ \tilde p_{\alpha_1}
\Psi^{\alpha_1}_{- i}(G_{eff}^{-1})^{ij}(\Psi_{eff})^{\beta_{1}}_j
p_{\beta_{1}}\right]+\frac{4}{\kappa}\Re \left[ \tilde
p_{\alpha_2} \Psi^{\alpha_2}_{+
i}(G_{eff}^{-1})^{ij}(\Psi_{eff})^{\beta_{2}}_j
p_{\beta_{2}}\right] \nonumber \\ &-&\frac{1}{\kappa}\Re\left[
p_{\alpha_1}\left(
 f_{14}^{\alpha_1\beta_{1}}+4\kappa
\Psi^{\alpha_1}_{- i}\Theta^{*i\beta_{1}}\right)
p^*_{\beta_{1}}\right]+\frac{1}{\kappa}\Re\left( \tilde p_{\alpha_1} f_{14}^{\alpha_1\beta_1}\tilde p^*_{\beta_1}\right)\nonumber \\
&+&\frac{4}{\kappa}\Re \left[ \tilde p_{\alpha_1}
\Psi^{\alpha_1}_{- i}(G_{eff}^{-1})^{ij}(\Psi_{eff})^{*\beta_{1}}_j p^*_{\beta_{1}}\right]\nonumber \\&-&\frac{1}{\kappa}\Re\left[ p_{\alpha_2}\left(
 f_{23}^{\alpha_2\beta_{2}}+4\kappa
\Psi^{\alpha_2}_{+ i}\Theta^{*i\beta_{2}}\right)
p^*_{\beta_{2}}\right]+\frac{1}{\kappa}\Re\left( \tilde p_{\alpha_2} f_{23}^{\alpha_2\beta_2}\tilde p^*_{\beta_2}\right)\nonumber \\
&+&\frac{4}{\kappa}\Re \left[ \tilde p_{\alpha_2}
\Psi^{\alpha_2}_{+
i}(G_{eff}^{-1})^{ij}(\Psi_{eff})^{*\beta_{2}}_j
p^*_{\beta_{2}}\right]\nonumber\\
&+&\frac{4}{\kappa}\Re\left[\tilde p_{\alpha_1}
\Psi^{\alpha_1}_{-i}(G_{eff}^{-1})^{ij}(\Psi_{eff})^{\beta_2}_j
p_{\beta_2}\right]+\frac{4}{\kappa}\Re\left[\tilde p_{\alpha_2}
\Psi^{\alpha_2}_{+ i}(G_{eff}^{-1})^{ij}(\Psi_{eff})^{\beta_1}_j
p_{\beta_1}\right]\nonumber \\
&-&\frac{1}{\kappa}\Re\left[  p_{\alpha_1}\left( f_{12}^{\alpha_1\beta_2}+ 4\kappa\Psi^{\alpha_1}_{-i}\Theta^{i\beta_2}\right) p_{\beta_2}\right] -\frac{1}{\kappa}\Re\left[ p_{\alpha_2}\left(f_{21}^{\alpha_2\beta_1}+4\kappa\Psi^{\alpha_2}_{+i}\Theta^{i\beta_1}\right)p_{\beta_1}\right] \nonumber
\\ &+&\frac{1}{\kappa}\Re\left(  \tilde p_{\alpha_1}f_{12}^{\alpha_1\beta_2} \tilde p_{\beta_2}\right) +\frac{1}{\kappa}\Re\left( \tilde p_{\alpha_2}f_{21}^{\alpha_2\beta_1}\tilde p_{\beta_1}\right) \nonumber
\\&+&\frac{4}{\kappa}\Re \left[ \tilde p_{\alpha_2}
\Psi^{\alpha_2}_{+
i}(G_{eff}^{-1})^{ij}(\Psi_{eff})^{*\beta_{2}}_j
p^*_{\beta_{2}}\right]+\frac{4}{\kappa}\Re\left[\tilde
p_{\alpha_2}
\Psi^{\alpha_2}_{+i}(G_{eff}^{-1})^{ij}(\Psi_{eff})^{*\beta_1}_j
p^*_{\beta_1}\right]\nonumber\\
&-&\frac{1}{\kappa}\Re\left[p_{\alpha_1}\left(f_{13}^{\alpha_1\beta_2}+4\kappa\Psi^{\alpha_1}_{-i}\Theta^{*i\beta_2}\right)p^*_{\beta_2}\right]-\frac{1}{\kappa}\Re\left[p_{\alpha_2}\left(f_{24}^{\alpha_2\beta_1}+4\kappa\Psi^{\alpha_2}_{+i}\Theta^{*i\beta_1}\right)p^*_{\beta_1}\right]\nonumber \\
&+&\frac{1}{\kappa}\Re\left(\tilde p_{\alpha_1} f_{13}^{\alpha_1\beta_2}\tilde p^*_{\beta_2}\right)+\frac{1}{\kappa}\Re\left(\tilde p_{\alpha_2} f_{24}^{\alpha_2\beta_1} \tilde p^*_{\beta_1}\right)\, ,
\end{eqnarray}
where
\begin{equation}
t_{\pm}=\tilde
t_{\pm}=\mp\frac{1}{4\kappa}(G_{eff}^{-1})^{ij}\tilde I_{\pm
i}\tilde I_{\pm j}\, .
\end{equation}
From effective Lagrangian
\begin{eqnarray}\label{eq:efektivniL}
{\mathcal L}^{eff}&=&\dot q^i p_i+2\Re\left(\dot \xi^{\alpha_1}
p_{\alpha_1}+\dot \xi^{\alpha_2} p_{\alpha_2}+\dot{\tilde
\xi}^{\alpha_1} \tilde p_{\alpha_1}+\dot{\tilde \xi}^{\alpha_2}
\tilde p_{\alpha_2}\right)-\tilde {\mathcal H}_c\, ,
\end{eqnarray}
and equations of motion for momentum $p_i$, we find the relation
\begin{equation}\label{eq:pmig}
p_i=\kappa G^{eff}_{ij}\dot q^j+4\Re\left[ (\Psi_{eff})^{\alpha_{1}}_i
p_{\alpha_1}+(\Psi_{eff})^{\alpha_{2}}_i
p_{\alpha_2} \right]\, ,
\end{equation}
which enables us to eliminate effective bosonic momentum $p_i$.
Putting this relation into expression for effective current
$\tilde I_{\pm i}$ (\ref{eq:efstruja}) we obtain
\begin{equation}\label{eq:strujaL}
\tilde I_{\pm i}=\kappa G^{eff}_{ij}(\dot q^j\pm
q'^j)+4\Pi_{\mp i}{}^j \Re\left[\Psi^{\alpha_1}_{- j}(p_{\alpha_1}\pm\tilde
p_{\alpha_1})+\Psi^{\alpha_2}_{+ j}(p_{\alpha_2}\pm\tilde
p_{\alpha_2})\right]\, .
\end{equation}
Substituting last two expression into effective Lagrangian
(\ref{eq:efektivniL}) we obtain its final form
\begin{eqnarray}\label{eq:efl1}
{\mathcal L}^{eff}&=&\frac{\kappa}{2}G^{eff}_{ij}\eta^{mn}
\partial_m q^i
\partial_n q^j+2\Re \left[ \left(\dot\xi^{\alpha_1}-\tilde\xi'^{\alpha_1}+2(\Psi_{eff})^{\alpha_{1}}_i\dot
q^i\right) p_{\alpha_1} \right]\nonumber \\  &+&2\Re    \left[
\left(\dot{\tilde\xi}^{\alpha_1}-\xi'^{\alpha_1}-2
(\Psi_{eff})^{\alpha_{1}}_i q'^i \right) \tilde
p_{\alpha_1}\right]\nonumber \\&+& 2\Re\left[
\left(\dot\xi^{\alpha_2}-{\tilde
\xi}'^{\alpha_2}+2(\Psi_{eff})^{\alpha_{2}}_i\dot
q^i\right)p_{\alpha_2} + \left(
\dot{\tilde\xi}^{\alpha_2}-\xi'^{\alpha_2}-2
(\Psi_{eff})^{\alpha_{2}}_i q'^i\right )  \tilde
p_{\alpha_2}\right]\nonumber  \\&+&\frac{1}{\kappa}\Re\left[
p_{\alpha_1}
 (f^{eff}_{11})^{\alpha_1\beta_{1}}
p_{\beta_{1}}- \tilde p_{\alpha_1}
(f_{11}^{eff})^{\alpha_1\beta_1}\tilde
p_{\beta_1}+p_{\alpha_2}(f^{eff}_{22})^{\alpha_2\beta_2}p_{\beta_2}-\tilde
p_{\alpha_2}(f^{eff}_{22})^{\alpha_2\beta_2}\tilde
p_{\beta_2}\right]\nonumber \\&+&\frac{2}{\kappa}\Re\left[p_{\alpha_1}(f_{12}^{eff})^{\alpha_1\beta_2} p_{\beta_2}-\tilde p_{\alpha_1}(f_{12}^{eff})^{\alpha_1\beta_2} \tilde p_{\beta_2}\right]\nonumber  \\
&+&\frac{1}{\kappa}\Re\left[
p_{\alpha_1}\bar f_{14}^{\alpha_1\beta_1} p^*_{\beta_1}-\tilde
p_{\alpha_1}\bar f_{14}^{\alpha_1\beta_1} \tilde
p^*_{\beta_1}+p_{\alpha_2}\bar f_{23}^{\alpha_2\beta_2} p^*_{\beta_2}-\tilde
p_{\alpha_2}\bar f_{23}^{\alpha_2\beta_2}\tilde
p^*_{\beta_2}\right]\nonumber \\
&+&\frac{2}{\kappa}\Re\left[p_{\alpha_1}\bar f_{13}^{\alpha_1\beta_2} p^*_{\beta_2}-\tilde p_{\alpha_1}\bar f_{13}^{\alpha_1\beta_2} \tilde p^*_{\beta_2}\right]\, ,
\end{eqnarray}
where effective background fields $G^{eff}_{ij}$,
$(\Psi_{eff})^{\alpha_{1}}_i$, $(\Psi_{eff})^{\alpha_{2}}_i$,
$(f^{eff}_{11})^{\alpha_1\beta_{1}}$, $(f_{12}^{eff})^{\alpha_1\beta_2}$ and
$(f^{eff}_{22})^{\alpha_2\beta_{2}}$ are defined in
Eq.(\ref{eq:effbackground}).
We also introduce notation for improved $D5$-brane bispinors
\begin{equation}\label{eq:polja1}
\bar f_{14}^{\alpha_1\beta_1}=\frac{1}{2}(f_{14}^{\alpha_1\beta_1}-f_{14}^{*\beta_1\alpha_1})-\Psi^{\alpha_1}_{-i}G^{ij}\Psi^{*\beta_1}_{-j}\, ,\quad \bar f_{23}^{\alpha_2\beta_2}=\frac{1}{2}(f_{23}^{\alpha_2\beta_2}-f_{23}^{*\beta_2\alpha_2})-\Psi^{\alpha_2}_{+ i}G^{ij}\Psi^{*\beta_2}_{+ j}\, ,
\end{equation}
\begin{equation}\label{eq:barf13}
\bar f_{13}^{\alpha_1\beta_2}=\frac{1}{2}\left(f_{13}^{\alpha_1\beta_2}-f_{24}^{*\beta_2\alpha_1}\right)-\Psi^{\alpha_1}_{-i}G^{ij}\Psi^{*\beta_2}_{+j}\, ,
\end{equation}
which have similar forms as effective ones (\ref{eq:effbackground}), but which do not appear in algebra of constraints (\ref{eq:algebraveza}).
Note that all terms in $\mathcal
L^{eff}$ are $\Omega$ even.

In the supersymmetric case $\Psi^{\alpha_1}_{- i}$ and
$\Psi^{\alpha_2}_{+i}$ play the same role as the antisymmetric
field $B_{ij}$ in pure bosonic case. In fact, none of them appears
explicitly, but they contribute as the bilinear terms in the
background fields of effective theory (\ref{eq:effbackground}),
(\ref{eq:polja1}) and (\ref{eq:barf13}).

Initial Lagrangian (\ref{eq:SB1}) depends on variables $x^i$,
$\theta^{\alpha_1}$, $\theta^{\alpha_2}$, $\bar\theta^{\alpha_1}$,
$\bar\theta^{\alpha_2}$, $\pi_{\alpha_1}$, $\pi_{\alpha_2}$,
$\bar\pi_{\alpha_1}$ and $\bar\pi_{\alpha_2}$, and effective
Lagrangian depends on the effective variables with well defined
$\Omega$ parity $q^i$, $\xi^{\alpha_1}$, $\xi^{\alpha_2}$, $\tilde
\xi^{\alpha_1}$, $\tilde\xi^{\alpha_2}$, $p_{\alpha_1}$,
$p_{\alpha_2}$, $\tilde p_{\alpha_1}$ and $\tilde p_{\alpha_2}$.
In order to compare initial and effective theories we have to make
correspondence between variables. If we substitute initial
variables $x^i$, $\theta^{\alpha_1}$, $\theta^{\alpha_2}$,
$\bar\theta^{\alpha_1}$ and $\bar\theta^{\alpha_2}$ with momenta
independent parts of their solutions, $x^i\to q^i$,
$\theta^{\alpha_1}\to \frac{1}{2}(\xi^{\alpha_1}+\tilde
\xi^{\alpha_1})$, $\theta^{\alpha_2}\to
\frac{1}{2}(\xi^{\alpha_2}+\tilde \xi^{\alpha_2})$,
$\bar\theta^{\alpha_1} \to \frac{1}{2}(\xi^{\alpha_1}-\tilde
\xi^{\alpha_1})$, $\bar\theta^{\alpha_2} \to
-\frac{1}{2}(\xi^{\alpha_2}-\tilde \xi^{\alpha_2})$, and fermionic
momenta with their complete solutions
(\ref{eq:resenje1})-(\ref{eq:resenje22}), $\pi_{\alpha_1}\to
p_{\alpha_1}+\tilde p_{\alpha_1}$, $\pi_{\alpha_2}\to
p_{\alpha_2}+\tilde p_{\alpha_2}$, $\bar\pi_{\alpha_1}\to
p_{\alpha_1}-\tilde p_{\alpha_1}$ and $\bar\pi_{\alpha_2}\to
-p_{\alpha_2}+\tilde p_{\alpha_2}$ in Lagrangian (\ref{eq:SB1}),
we obtain
\begin{eqnarray}\label{eq:eflp}
{\mathcal L}&\to&\frac{\kappa}{2}G_{ij}\eta^{mn}
\partial_m q^i
\partial_n q^j+\kappa\varepsilon^{mn}B_{ij}\partial_m q^i
\partial_n q^j\nonumber \\&+&2\Re \left[ \left(\dot\xi^{\alpha_1}-\tilde\xi'^{\alpha_1}+\Psi^{\alpha_1}_{+i}\dot
q^i-\Psi^{\alpha_1}_{-i}q'^i\right) p_{\alpha_1} +
\left(\dot{\tilde\xi}^{\alpha_1}-\xi'^{\alpha_1}+\Psi^{\alpha_1}_{-i}\dot q^i-\Psi^{\alpha_{1}}_{+i} q'^i \right) \tilde
p_{\alpha_1}\right]\nonumber \\&+&2\Re \left[ \left(\dot\xi^{\alpha_2}-\tilde\xi'^{\alpha_2}+\Psi^{\alpha_2}_{-i}\dot
q^i-\Psi^{\alpha_2}_{+i}q'^i\right) p_{\alpha_2} +
\left(\dot{\tilde\xi}^{\alpha_2}-\xi'^{\alpha_2}+\Psi^{\alpha_2}_{+i}\dot q^i-\Psi^{\alpha_{2}}_{-i} q'^i \right) \tilde
p_{\alpha_2}\right]\nonumber  \\&+&\frac{1}{\kappa}\Re\left[
p_{\alpha_1}
 (f^{a}_{11})^{\alpha_1\beta_{1}}
p_{\beta_{11}}- \tilde p_{\alpha_1}
(f_{11}^{a})^{\alpha_1\beta_1}\tilde
p_{\beta_1}+p_{\alpha_2}(f^{a}_{22})^{\alpha_2\beta_2}p_{\beta_2}-\tilde
p_{\alpha_2}(f^{a}_{22})^{\alpha_2\beta_2}\tilde
p_{\beta_2}\right]\nonumber \\&+&\frac{2}{\kappa}\Re\left[
\tilde p_{\alpha_1}
 (f^{s}_{11})^{\alpha_1\beta_{1}}
p_{\beta_{1}}+ \tilde p_{\alpha_2}
 (f^{s}_{22})^{\alpha_2\beta_{2}}
p_{\beta_{2}}\right]\nonumber \\&+&\frac{2}{\kappa}\Re\left[p_{\alpha_1}\frac{1}{2}(f_{12}^{\alpha_1\beta_2}-f_{21}^{\beta_2\alpha_1}) p_{\beta_2}-\tilde p_{\alpha_1}\frac{1}{2}(f_{12}^{\alpha_1\beta_2}-f_{21}^{\beta_2\alpha_1}) \tilde p_{\beta_2}\right]\nonumber  \\&+&\frac{1}{\kappa}\Re\left[\tilde p_{\alpha_1}(f_{12}^{\alpha_1\beta_2}+f_{21}^{\beta_2\alpha_1}) p_{\beta_2}-p_{\alpha_1}(f_{12}^{\alpha_1\beta_2}+f_{21}^{\beta_2\alpha_1}) \tilde p_{\beta_2}\right]\nonumber  \\
&+&\frac{1}{\kappa}\Re\left[
p_{\alpha_1} \frac{1}{2}(f_{14}^{\alpha_1\beta_1}-f_{14}^{*\beta_1\alpha_1}) p^*_{\beta_1}-\tilde
p_{\alpha_1} \frac{1}{2}(f_{14}^{\alpha_1\beta_1}-f_{14}^{*\beta_1\alpha_1}) \tilde
p^*_{\beta_1}\right]\nonumber \\&+&\frac{1}{\kappa}\Re\left[p_{\alpha_2}\frac{1}{2}(f_{23}^{\alpha_2\beta_2}-f_{23}^{*\beta_2\alpha_2}) p^*_{\beta_2}-\tilde
p_{\alpha_2}\frac{1}{2}(f_{23}^{\alpha_2\beta_2}-f_{23}^{*\beta_2\alpha_2})\tilde
p^*_{\beta_2}\right]\nonumber \\&-&\frac{1}{\kappa}\Re\left[
p_{\alpha_1} (f_{14}^{\alpha_1\beta_1}+f_{14}^{*\beta_1\alpha_1}) \tilde p^*_{\beta_1}+p_{\alpha_2} (f_{23}^{\alpha_2\beta_2}+f_{23}^{*\beta_2\alpha_2}) \tilde p^*_{\beta_2}\right]\nonumber \\
&+&\frac{2}{\kappa}\Re\left[p_{\alpha_1}\frac{1}{2}(f_{13}^{\alpha_1\beta_2}-f_{24}^{*\beta_2\alpha_1}) p^*_{\beta_2}-\tilde p_{\alpha_1}\frac{1}{2}(f_{13}^{\alpha_1\beta_2}-f_{24}^{*\beta_2\alpha_1}) \tilde p^*_{\beta_2}\right]\nonumber \\
&+&\frac{1}{\kappa}\Re\left[\tilde p_{\alpha_1}(f_{13}^{\alpha_1\beta_2}+f_{24}^{*\beta_2\alpha_1}) p^*_{\beta_2}-p_{\alpha_1}(f_{13}^{\alpha_1\beta_2}+f_{24}^{*\beta_2\alpha_1}) \tilde p^*_{\beta_2}\right]\, .
\end{eqnarray}
Comparing this Lagrangian with effective one (\ref{eq:efl1}) we can
conclude that $ \mathcal L \to \mathcal L^{eff}$ after substitution
\begin{eqnarray}\label{eq:prelaz}
&{}& G_{ij}\to G_{ij}^{eff}\, ,\quad\qquad\qquad\qquad\qquad\qquad\qquad\qquad\quad\quad\; B_{ij}\to 0\, ,\nonumber \\
&{}& \frac{1}{2}\Psi^{\alpha_1}_{+i}\to (\Psi_{eff})^{\alpha_1}_i\, ,\quad \frac{1}{2}\Psi^{\alpha_2}_{-i}\to (\Psi_{eff})^{\alpha_2}_i\, ,\quad\quad\quad\quad\quad\; \Psi^{\alpha_1}_{-i}\to 0\, ,\quad \Psi^{\alpha_2}_{+i}\to 0\, ,\nonumber \\
&{}& (f^a_{11})^{\alpha_1\beta_1}\to (f^{eff}_{11})^{\alpha_1\beta_1} \, ,\,  (f^a_{22})^{\alpha_2\beta_2}\to (f^{eff}_{22})^{\alpha_2\beta_2} \, ,\quad\quad (f^s_{11})^{\alpha_1\beta_1}\to 0\, ,\quad (f^s_{22})^{\alpha_2\beta_2}\to 0\, , \nonumber \\
&{}& \frac{1}{2}(f_{12}^{\alpha_1\beta_2}-f_{21}^{\beta_2\alpha_1})\to (f_{12}^{eff})^{\alpha_1\beta_2}\, ,\quad\quad\qquad\qquad\qquad\quad f_{12}^{\alpha_1\beta_2}+f_{21}^{\beta_2\alpha_1}\to 0\, ,\nonumber \\
&{}& \frac{1}{2}(f_{14}^{\alpha_1\beta_1}-f_{14}^{*\beta_1\alpha_1})\to \bar f_{14}^{\alpha_1\beta_1}\, ,\; \, \quad\quad\quad\quad\quad\qquad\qquad\quad f_{14}^{\alpha_1\beta_1}+f_{14}^{*\beta_1\alpha_1}\to 0\, \; \, ,\nonumber \\&{}& \frac{1}{2}(f_{23}^{\alpha_2\beta_2}-f_{23}^{*\beta_2\alpha_2})\to \bar f_{23}^{\alpha_2\beta_2}\, ,\; \, \quad\quad\quad\quad\quad\qquad\qquad\quad f_{23}^{\alpha_2\beta_2}+f_{23}^{*\beta_2\alpha_2}\to 0\nonumber \\
&{}& \frac{1}{2}(f_{13}^{\alpha_1\beta_2}-f_{24}^{*\beta_2\alpha_1})\to \bar f_{13}^{\alpha_1\beta_2}\, ,\qquad\qquad\qquad\quad\quad\quad\qquad f_{13}^{\alpha_1\beta_2}+f_{24}^{*\beta_2\alpha_1}\to 0\, .
\end{eqnarray}

In our approach the effective theory has been obtained from
initial IIB one with embedded $D5$-brane on the solution of
boundary conditions. The condition of $2\pi$ periodicity of all
canonical variables solves boundary condition at $\sigma=\pi$ in
terms of that at $\sigma=0$, and consequently, this is a closed
string theory. The effective theory is $\Omega$ symmetric part of
type IIB superstring theory, and we expect that it should
correspond to the type I superstring theory with embedded
$D5$-brane, which will be shown in the next subsection.

\subsection{Type I superstring theory with $D5$-brane}

In order to work with stable initial and final theories we should embed
$D5$-brane both in type IIB and type I string theories. We have
already done this for type IIB theory and now we will apply
similar procedure for type I superstring theory. The corresponding
Lagrangian is of the form \cite{BNBSPLB}
$$\mathcal L^{I}=
\frac{\kappa}{2}G^{I}_{\mu\nu}\eta^{mn}\partial_m q^\mu
\partial_n q^\nu+ \hphantom{mmmmmmmmmmmmmmmmmmm} $$
\begin{equation}\label{eq:xxx}
 -\pi_\alpha
(\partial_\tau-\partial_\sigma)\left[ \eta^\alpha+(\Psi_{I})^\alpha_\mu
q^\mu\right] +( \partial_\tau+\partial_\sigma)\left[ \bar\eta^{\alpha}+
(\Psi_{I})^{\alpha}_\mu q^\mu\right] \bar \pi_{\alpha} -\frac{1}{2\kappa} \pi_\alpha (F_{I}{}^\star \Gamma)^{\alpha
\beta}\bar \pi_{\beta} \, .
\end{equation}
Index $I$ stands for type I superstring theory and variable
$q^\mu$ is symmetric under $\Omega$ transformation. The field
strength $(F_{I}{}^\star \Gamma)^{\alpha\beta}$ is antisymmetric
under permutation of indices, $(F_{I}{}^\star
\Gamma)^{\alpha\beta}=-(F_{I}{}^\star \Gamma)^{\beta\alpha}$,
where matrix ${}^\star \Gamma$ is defined as (see Appendix A)
\begin{equation}\label{eq:gamazv}
{}^\star \Gamma\equiv \Gamma^0\Gamma^1\Gamma^2\Gamma^3\Gamma^4\Gamma^5=\left(\begin{array}{cccc}
-\gamma & 0 & 0 & 0\\
0 & -\gamma & 0 & 0\\
0 & 0 & -\gamma & 0\\
0 & 0 & 0 & -\gamma
\end{array}\right)\, .
\end{equation}

Let us embed $D5$-brane in this theory. For coordinates $q^i$
($i=0,1,\dots,5$) we choose Neumann  and for $q^a$ ($a=6,7,8,9$)
Dirichlet boundary conditions. We can take that $q^i$ and $q^a$
are orthogonal coordinates which implies $G^{I}_{ia}=0$. Also we
take that field $(\Psi_I)^\alpha_\mu$ is
nontrivial only on $D5$-brane, $(\Psi_I)^\alpha_\mu \to (\Psi_I)^\alpha_i$. The term which
describes the free string oscillation in $q^a$ directions decouples
from the rest.

In Table 2 we
summarize the list of the background fields of the effective
theory in $D=10$ dimensional space-time, fields living on the
$D5$-brane and the rest fields.

\begin{table}[h]
\begin{tabular}{|c|c|c|c|}\hline
Sector  & $x^\mu\,(\mu=0,1,\dots,9)$    & $x^i\,(i=0,1,\dots,5)$ & $x^a\, (a=6,7,8,9)$ \\ \hline\hline
\raisebox{-1.5ex}[0pt]{NS-NS} & $G^{eff}_{\mu\nu}$ & $G^{eff}_{ij}$ & $G^{eff}_{ab}$ (decoupled) \\
            & $\Phi=0$ &  & \\ \hline
\raisebox{0.0ex}[0pt]{NS-R} & $(\Psi_{eff})^\alpha_\mu$ & $(\Psi_{eff})^{\alpha_1}_i\, ,(\Psi_{eff})^{\alpha_2}_i$ &  \\ \hline
\raisebox{-1.5ex}[0pt]{R-R} & \raisebox{-1.5ex}[0pt]{$F_{eff}^{\alpha\beta}$} & $(f^{eff}_{11})^{\alpha_1\beta_1}, (f_{22}^{eff})^{\alpha_2\beta_2}, (f_{12}^{eff})^{\alpha_1\beta_2}$ & \\
                            &  & $\bar f_{14}^{\alpha_1\beta_1}, \bar f_{23}^{\alpha_2\beta_2}, \bar f_{13}^{\alpha_1\beta_2}$ & \\ \hline
\end{tabular}
\caption{Background fields of the effective theory: the complete set, part living on $D5$-brane and the rest fields eliminated from the theory}
\end{table}

Now, using the relation (\ref{eq:relacija1}) and (\ref{eq:pFp}), we rewrite the
Lagrangian (\ref{eq:xxx}) in terms of $D5$-brane variables and background fields. We want
to establish relation of such theory with effective one
(\ref{eq:efl1}). Comparing terms linear in fermionic momenta and
independent of background fields, we obtain the following
connection between coordinates and momenta of these theories
\begin{eqnarray}
\eta^\alpha&=&\frac{1}{2}(\xi^\alpha+\tilde\xi^\alpha)\, ,\quad \bar\eta^\alpha=-\frac{1}{2}\left[{}^\star\Gamma(\xi-\tilde\xi)\right]^\alpha\, ,\nonumber \\
\pi_\alpha&=&p_\alpha+\tilde p_\alpha\, ,\quad \bar\pi_\alpha=-\left[{}^\star\Gamma(p-\tilde p)\right]_\alpha\, ,
\end{eqnarray}
where
\begin{equation}
\xi^\alpha=P_s \theta^\alpha=-P_s({}^\star \Gamma \bar\theta)^\alpha\, ,\quad \tilde\xi^\alpha=P_a\left[\theta^\alpha+({}^\star \Gamma\bar\theta)^\alpha\right]\, ,\nonumber
\end{equation}
\begin{equation}\label{eq:varijable10}
p_\alpha=P_s \pi_\alpha=-P_s (\bar\pi {}^\star \Gamma)_\alpha\, ,\quad \tilde p_\alpha=P_a \pi_\alpha=P_a(\bar\pi{}^\star \Gamma)_\alpha\, .
\end{equation}
Note that these are just relations (\ref{eq:resenjepi}) rewritten
in terms of ten dimensional spinors. Putting these relations into
expression for $\mathcal L^{I}$, we have
\begin{eqnarray}\label{eq:efl3}
{\mathcal L}^{I}(\mathcal A^I)=\mathcal L^{eff}(\mathcal A^{eff})\, ,
\end{eqnarray}
where $\mathcal A^I$ and $\mathcal A^{eff}$ denote background
fields of type I superstring theory and the effective theory,
respectively. So, $D5$-brane background fields of type I theory
can be expressed in terms of the corresponding ones of type IIB
theory as
\begin{equation}\label{eq:poc}
G^I_{ij}=G_{ij}^{eff}\, ,\quad (\Psi_I)^{\alpha_1}_i=(\Psi_{eff})^{\alpha_1}_i\, ,\quad (\Psi_I)^{\alpha_2}_i=(\Psi_{eff})^{\alpha_2}_i\, ,\nonumber
\end{equation}
\begin{equation}
(f^{a\,(I)}_{11})^{\alpha_1\beta_{1}}=(f^{eff}_{11})^{\alpha_1\beta_1}\, ,\quad (f^{a\,(I)}_{22})^{\alpha_2\beta_{2}}=(f^{eff}_{22})^{\alpha_2\beta_2}\, ,\nonumber
\end{equation}
\begin{equation}\label{eq:IprekoII}
\frac{1}{2}(f_{12}^{(I)\,\alpha_1\beta_2}-f_{21}^{(I)\,\beta_2\alpha_1})=(f_{12}^{eff})^{\alpha_1\beta_2}\, ,
\end{equation}
\begin{eqnarray}\label{eq:kraj}
\frac{1}{2}(f_{14}^{(I)\,\alpha_1\beta_1}-f_{14}^{*(I)\,\beta_1\alpha_1})=\bar f_{14}^{\alpha_1\beta_1}\, &,&\quad \frac{1}{2}(f_{23}^{(I)\,\alpha_2\beta_2}-f_{23}^{*(I)\,\beta_2\alpha_2})=\bar f_{23}^{\alpha_2\beta_2}\, ,\nonumber \\ \frac{1}{2}(f_{13}^{(I)\,\alpha_1\beta_2}&-&f_{24}^{*(I)\,\beta_2\alpha_1})=\bar f_{13}^{\alpha_1\beta_2}\, ,\nonumber
\end{eqnarray}
where the right-hand sides are defined in
(\ref{eq:effbackground}), (\ref{eq:polja1}) and (\ref{eq:barf13}).
The bispinors from the last three lines in Eq.(\ref{eq:IprekoII})
can be written in terms of antisymmetric tensors (see Appendix B)
\begin{eqnarray}
&{}&f_{(0)}^{(I)}=f_{(0)}-\frac{1}{6}G^{ij}\Psi_{\pm i}\Psi_{\pm j}\, , \quad f_{ij}^{(I)}=f_{ij}-\frac{1}{6}G^{kl}\Psi_{\pm k}\gamma_{[ij]}\Psi_{\pm l}\, ,\nonumber\\
&{}&f_i^{(I)}=f_i-\frac{1}{6}G^{kl}\Psi_{\pm k}\gamma_i \Psi_{\pm l}\, ,\quad f_{ijk}^{(I)}=f_{ijk}-\frac{1}{6}G^{lm}\Psi_{\pm l}\gamma_{[ijk]} \Psi_{\pm m}\, .
\end{eqnarray}
Here we write general forms of type I $D5$-brane R-R background fields in terms of
type IIB ones. Field strength $f_i$
and $f_{ijk}$ appear as the coefficients in gamma matrix expansion
of the bispinors $f_{11}$, $f_{22}$, $f_{14}$ and $f_{23}$, while
$f_{(0)}$ and $f_{ij}$ in the same sense are related to $f_{12}$,
$f_{21}$, $f_{13}$ and $f_{24}$. Because $f^{a\,(I)}_{11}$ and
$f^{a\,(I)}_{22}$ are antisymmetric tensors, they can be expressed
in terms of $f_i$, while $f_{14}^{(I)}$ and $f_{23}^{(I)}$ contain
both $f_{i}$ and $f_{ijk}$ tensors.

Fields $B_{ij}$, $\Psi^{\alpha_1}_{-i}$ and
$\Psi^{\alpha_2}_{+i}$, odd under $\Omega$ parity transformation,
are not completely eliminated. They appear as bilinear terms in
type I $D5$-brane background fields. Consequently, we obtained the
generalized expressions for type I $D5$-brane background fields in
terms of the type IIB $D5$-brane background fields. The quadratic
terms in the expressions for effective background fields can be
considered as supersymmetric generalization of the open string
metric $G^{eff}_{\mu\nu}$ obtained by Seiberg and Witten
\cite{SW}.

\subsection{Embedding of $D5$-brane in type IIB and type I superstring theory using T-duality}

Using the expressions for type I superstring background fields in
terms of type IIB ones obtained in \cite{BNBSPLB} and T-duality
transformations \cite{JT} along $x^a (a=6,7,8,9)$ directions, we
find the expressions for background fields obtained in previous
subsection.

In the third article of \cite{JT} type IIA/B superstring theory is
considered and finding of T-duality transformation along one
direction is demonstrated. Here we will apply the presented
procedure for type IIB superstring theory and four $x^a$
directions which are chosen to be orthogonal to $D5$-brane.

We start with the action of type IIB superstring theory in pure spinor formulation given in (\ref{eq:SB}) in the form
\begin{eqnarray}\label{eq:SBT}
&{}&S=\kappa \int_\Sigma d^2\xi \partial_{+}x^\mu \Pi_{+\mu\nu}\partial_- x^\nu \\&+&\int_\Sigma d^2 \xi \left[ -\pi_\alpha
\partial_-(\theta^\alpha+\Psi^\alpha_\mu
x^\mu)+\partial_+(\bar\theta^{\alpha}+\bar
\Psi^{\alpha}_\mu x^\mu)\bar\pi_{\alpha}+\frac{1}{2\kappa}\pi_\alpha F^{\alpha
\beta}\bar \pi_{\beta}\right ]\, ,\nonumber
\end{eqnarray}
where the background fields are constant, $\Pi_{\pm
\mu\nu}=B_{\mu\nu}\pm\frac{1}{2}G_{\mu\nu}$ and
$\partial_{\pm}=\partial_\tau \pm \partial_\sigma$. We suppose that the action has a
global shift symmetry in $x^a$ ($a=6,7,8,9$) directions. So, we have to introduce
gauge fields $v^a_{\pm}$ and make a change in the action
\begin{equation}
\partial_{\pm}x^a \to \partial_{\pm}x^a+v_{\pm}^a\, .
\end{equation}
For the fields $v_\pm^a$ we have to introduce additional term in the action
\begin{equation}
S_{gauge}=\frac{1}{2}\kappa \int_\Sigma d^2\xi y_a (\partial_+ v_-^a-\partial_- v^a_+)\, ,
\end{equation}
which produces vanishing of the field strength $\partial_+
v_-^a-\partial_- v^a_+$ if we vary the action with respect to the
Lagrange multipliers $y_a$. The full action has the form
\begin{equation}
S^\star=S(x^i,x^a,v^a_\pm)+S_{gauge}(y^a,v^a_\pm)\, .
\end{equation}
Let us note that on the equations of motion for $y_a$ we have
$v^a_\pm=\partial_\pm x^a$ and the original dynamics survives
unchanged.

Now we can fix $x^a$ to zero and obtain the action quadratic in
the fields $v_\pm^a$, which can be integrated out classically. On
the equations of motion for $v^a_\pm$ we obtain expressions for
these gauge fields in terms of $x^i$ ($i=0,1,\dots,5$), $y_a$ and
momenta, $\pi_\alpha$ and $\bar\pi_\alpha$,
\begin{eqnarray}
&{}& v^a_+=2(2\partial_+ x^i \Pi_{+ib}-\partial_+y_b-\frac{2}{\kappa}\pi_\alpha \Psi^\alpha_b)\Theta_-^{ba}\, ,\\
&{}& v^a_-=2\Theta_-^{ab}(2\Pi_{+bi}\partial_- x^i+\partial_-y_b+\frac{2}{\kappa}\bar\Psi^\alpha_b \bar\pi_\alpha)\, ,
\end{eqnarray}
where
\begin{equation}
\Theta^{ab}_-=(G^{-1}\Pi_-g_{eff}^{-1})^{ab}\, ,\quad g^{eff}_{ab}\equiv -4 \Pi_{\pm ac}G^{cd}\Pi_{\mp db}=G_{ab}-4B_{ac}G^{cd}B_{db}\, .
\end{equation}

Substituting expression for $v^a_\pm$ in the action $S^{\star}$ we
obtain the the dual action from which we read the dual background
fields
\begin{eqnarray}
&{}& \tilde\Pi_{+ij}=\Pi_{+ij}+4 \Pi_{+ ia}\Theta_-^{ab}\Pi_{+bj}\, ,\label{eq:1}\\ &{}& \tilde\Pi_{+i}{}^a=2\Pi_{+ib}\Theta_-^{ba}\, ,\quad \tilde \Pi^a{}_{+ i}=-2\Theta_{-}^{ab}\Pi_{+bi}\, ,\label{eq:2}\\  &{}& \tilde\Pi_+^{ab}=-\Theta_-^{ab}\, ,\label{eq:3}\\
&{}& \tilde\Psi^{\alpha}_i= \Psi^\alpha_i+4 \Psi^\alpha_a \Theta_-^{ab}\Pi_{+b i}\, ,\\ &{}& \tilde{\bar\Psi}^{\alpha}_i=-({}^\star \Gamma\bar\Psi)^\alpha_i-4\Pi_{+ia}\Theta_-^{ab}({}^\star\Gamma\bar\Psi)^\alpha_b\, , \\ &{}& \tilde\Psi^{\alpha a}=2 \Psi^\alpha_b \Theta_-^{ba}\, ,\\ &{}& \tilde{\bar\Psi}^{\alpha a}=2 \Theta_-^{ab} ({}^\star\Gamma\bar\Psi)^{\alpha}_b\, ,\\ &{}& \tilde F^{\alpha\beta}=-\left[(F-8\Psi_a \Theta_-^{ab}\bar\Psi_b){}^\star\Gamma\right]^{\alpha\beta}\, .
\end{eqnarray}
Symbol $\tilde A$ denotes dual background field of the arbitrary field $A$. We redefine the fields $\bar\Psi^\alpha_\mu$ and $F^{\alpha\beta}$ as well as all other bar variables multiplying them with $-{}^\star\Gamma$ (\ref{eq:gamazv}). From (\ref{eq:1})-(\ref{eq:3}) we obtain the T-duality transformation rules for the fields $G_{\mu\nu}$ and $B_{\mu\nu}$
\begin{eqnarray}
\tilde G_{ij}&=&G_{ij}+4G_{ia}\Theta^{ab}B_{bj}+4B_{ia}\Theta^{ab}G_{bj}-4B_{ia}(g_{eff}^{-1})^{ab}B_{bj}-G_{ia}(g_{eff}^{-1})^{ab}G_{bj}\, ,\nonumber \\ \tilde G_i{}^{a}&=& 2 G_{ib} \Theta^{ba}-2B_{ib}(g_{eff}^{-1})^{ba}\, ,\quad \tilde G^{ab}=(g_{eff}^{-1})^{ab}\, ,
\end{eqnarray}
\begin{eqnarray}
\tilde B_{ij}&=&B_{ij}+4B_{ia}\Theta^{ab}B_{bj}+G_{ia}\Theta^{ab}G_{bj}-G_{ia}\Theta^{ab}B_{bj}-B_{ia}(g_{eff}^{-1})^{ab}G_{bj}\, ,\nonumber \\ \tilde B_i{}^{a}&=&2B_{ib}\Theta^{ba}-\frac{1}{2}G_{ib}(g^{-1}_{eff})^{ba}\, ,\quad \tilde B^{ab}= -\Theta^{ab}\, ,
\end{eqnarray}
where $\Theta^{ab}=(g_{eff}^{-1}BG^{-1})^{ab}$.

Our choice of background fields introduced in Section 2, $G_{ia}=0$, $B_{ia}=B_{ab}=0$, $\Psi^\alpha_a=\bar\Psi^\alpha_a=0$, implies
\begin{eqnarray}
&{}& \tilde G_{ij}=G_{ij}\, ,\quad \tilde G_{i}{}^{a}=0\, ,\quad \tilde G^{ab}=G^{ab}\, ,\\
&{}& \tilde B_{ij}=B_{ij}\, ,\quad \tilde B_{i}{}^{a}=0\, ,\quad \tilde B^{ab}=0\, ,\\
&{}& \tilde\Psi^{\alpha_1}_i=\Psi^{\alpha_1}_i\, ,\tilde\Psi^{\alpha_2}_i=\Psi^{\alpha_2}_i\, , \quad \tilde{\bar\Psi}^{\alpha_1}_i=\bar\Psi^{\star\alpha_1}_i\, ,\tilde{\bar\Psi}^{\alpha_2}_i=-\bar\Psi^{\alpha_2}_i\, ,\\
&{}& \tilde f_{11}^{\alpha_1\beta_1}=f_{11}^{\alpha_1\beta_1}\, ,\quad \tilde f_{12}^{\alpha_1\beta_2}=-f_{12}^{\alpha_1\beta_2}\, ,\quad \tilde f_{21}^{\alpha_2\beta_1}=f_{21}^{\alpha_2\beta_1}\, ,\quad \tilde f_{22}^{\alpha_2\beta_2}=-f_{22}^{\alpha_2\beta_2}\, , \\ &{}& \tilde f_{13}^{\alpha_1\beta_2}=-f_{13}^{\alpha_1\beta_2}\, ,\quad \tilde f_{14}^{\alpha_1\beta_1}=f_{14}^{\alpha_1\beta_1}\, ,\quad \tilde f_{23}^{\alpha_2\beta_2}=-f_{23}^{\alpha_2\beta_2}\, ,\quad \tilde f_{24}^{\alpha_2\beta_1}=f_{24}^{\alpha_2\beta_1}\, .
\end{eqnarray}

The same procedure we repeat for type I superstring theory which Lagrangian is given in (\ref{eq:xxx}).
The form of the dual field for type I superstring theory is the same as for type IIB one up to the condition $B_{\mu\nu}=0$ and $\Psi^\alpha_\mu=\bar\Psi^\alpha_\mu$. We have
\begin{equation}
\tilde G_{ij}^{I}=G_{ij}^{I}-G^{I}_{ia}G_{I}^{ab}G^{I}_{bj}\, ,\quad (\tilde G^{I})_i{}^a=0\, ,\quad \tilde G_{I}^{ab}=G_{I}^{ab}\, ,
\end{equation}
\begin{equation}
\tilde B_{ij}^{I}=0\, ,\quad (\tilde B^{I})_i{}^a=-\frac{1}{2}G^{I}_{ib}G_{I}^{ba}\, ,\quad \tilde B_{I}^{ab}=0\, ,
\end{equation}
\begin{equation}
(\tilde\Psi_{I})^\alpha_i=(\Psi_{I})^\alpha_i-(\Psi_{I})^\alpha_a (G_{I})^{ab} G^{I}_{bi}\, ,
\end{equation}
\begin{equation}
(\tilde\Psi_{I})^{\alpha a}=-(\Psi_{I})^\alpha_b G^{ba}_{I}\, ,
\end{equation}
\begin{equation}
\tilde F_{I}^{\alpha\beta}=F_{I}^{\alpha\beta}+4(\Psi_{I})^\alpha_a G_{I}^{ab} (\Psi_{I})^\alpha_b\, ,
\end{equation}
where for simlicity $F_{I}^{\alpha\beta}$ stands instead $-(F_I{}^\star\Gamma)^{\alpha\beta}$ used in Eq.(\ref{eq:xxx}).
Choosing background fields as in Section 2 of the present paper, we have
\begin{equation}
\tilde G^I_{ij}=G^I_{ij}\, ,\quad (\tilde G^I)_i{}^a=0\, ,\quad \tilde G_I^{ab}=G_I^{ab}\, ,
\end{equation}
\begin{equation}
(\tilde \Psi_I)^\alpha_\mu=(\Psi_I)^\alpha_\mu\, , \quad \tilde F_I^{\alpha\beta}=F_I^{\alpha\beta}\, ,
\end{equation}
\begin{equation}
\tilde B^{I}_{ij}=(\tilde B_I)_i{}^a=(\tilde B_I)^{ab}=0 \, ,\quad (\tilde{\bar\Psi})^{\alpha a}=0\, .
\end{equation}

Using obtained duality transformation for type IIB and type I superstring theories we can reproduct the results of the prsent paper from the results of Ref.\cite{BNBSPLB}. For example, from the expression for type I superstring metric obtained in \cite{BNBSPLB}
\begin{equation}
G^I_{\mu\nu}=G_{\mu\nu}-4B_{\mu\rho}G^{\rho\lambda}B_{\lambda\nu}\, ,
\end{equation}
after T-duality transformation we have
\begin{equation}
\tilde G^{I}_{ij}=\tilde G_{ij}-4\tilde B_{ik}\tilde G^{kl}\tilde B_{lj}-4 \tilde B_i{}^a \tilde B_{aj}=G^{eff}_{ij}=G^{I}_{ij}\, .
\end{equation}

\section{Concluding remarks}
\setcounter{equation}{0}

In this paper we considered relation between type IIB and type I
theories with embedded $D5$-branes. We used the pure spinor
formulation of the theories introduced in Refs.\cite{NPBref}
restricting our analysis to the quadratic terms. We suppose that
all background fields, the metric tensor $G_{\mu\nu}$,
antisymmetric NS-NS field $B_{\mu\nu}$, gravitino fields
$\Psi^\alpha_\mu$ and $\bar\Psi^{\alpha}_\mu$, and the R-R field
strength $F^{\alpha\beta}$, are constant.

In Ref.\cite{BNBSPLB} we showed that type IIB superstring theory,
on the solution of appropriately chosen open string boundary
conditions, corresponds to the type I superstring theory. It means
that we obtained relation between $D9$-branes in these theories.
In the present article, instead $D9$-brane we considered
$D5$-branes which are stable in both theories.

Using canonical method, following \cite{radepjc}, we derived
boundary conditions from the requirement that Hamiltonian, as time
translation generator, has well defined functional derivatives in
supercoordinates and their canonically conjugated supermomenta.
All boundary conditions at string endpoints we treated as
canonical constraints. Applying Dirac consistency procedure they
produced an infinite set of constraints. With the help of Taylor expansion they can be rewritten as five $\sigma$-dependent constraints.
All these constraints, originating from boundary conditions, are
of the second class and we can solve them. We obtained the
expressions for coordinates $x^i$, $\theta^{\alpha_1}$,
$\bar\theta^{\alpha_1}$, $\theta^{\alpha_2}$ and
$\bar\theta^{\alpha_2}$ in terms of effective ones, $q^i$,
$\xi^{\alpha_1}$, $\tilde\xi^{\alpha_1}$, $\xi^{\alpha_2}$
and $\tilde\xi^{\alpha_2}$ (momenta independent parts of the
solutions for initial supercoordinates $x^i$, $\theta^{\alpha_1}$,
$\bar\theta^{\alpha_1}$, $\theta^{\alpha_2}$ and
$\bar\theta^{\alpha_2}$) and momenta $p_i$, $p_{\alpha_1}$ and $p_{\alpha_2}$ (canonically conjugated to $q^i$, $\xi^{\alpha_1}$ and $\xi^{\alpha_2}$ respectively).

Effective Lagrangian, obtained on the solution of boundary
conditions, is even under orientifold projection $\Omega$ in six
dimensions. In fact it has form of type I superstring theory with
embedded $D5$-brane. As a result we obtained the expressions for
$D5$-brane background fields of type I theory in terms of the
corresponding ones of type IIB (\ref{eq:IprekoII}). Note that
second parts of effective backgrounds in (\ref{eq:effbackground}),
(\ref{eq:polja1}) and (\ref{eq:barf13}), bilinear in $\Omega$ odd
fields are our improvements to the well known first parts, linear
in $\Omega$ even fields.Seiberg and Witten \cite{SW} obtained term
with square of the Kalb-Ramond field $B_{\mu\nu}$ in the open
string metric $G_{\mu\nu}^{eff}$. Our quadratic parts of the
effective background fields can be considered as a supersymmetric
generalization of their result.

In subsection 5.3 we showed that there is a relation between the
results of the present paper and those from \cite{BNBSPLB}.
This connection is realized by T-duality transformations
along the $x^a$ directions, which are orthogonal to $D5$-brane.

Table 3 contains the background fields of the type IIB superstring
theory with embedded $D5$-brane, $\Omega$ even projection of type
IIB with $D5$-brane, effective theory and type I with $D5$-brane.

\begin{table}[h]
\begin{tabular}{|c|c|c|c|}\hline
Theory & NS-NS & NS-R &
R-R \\ \hline\hline
\raisebox{0.0ex}[0pt]{Type IIB} & $G_{ij}\, ,B_{ij}\, ,\Phi(=0)$ & $\Psi^{\alpha_1}_i$, $\Psi^{\alpha_2}_i$  & $f_{11}^{\alpha_1\beta_1}, f_{12}^{\alpha_1\beta_2}, f_{21}^{\alpha_2\beta_1}, f_{22}^{\alpha_2\beta_2}$ \\ & & $\bar\Psi^{\alpha_1}_i$, $\bar\Psi^{\alpha_2}_i$ & $f_{14}^{\alpha_1\beta_1}, f_{23}^{\alpha_2\beta_2}, f_{13}^{\alpha_1\beta_2}, f_{24}^{\alpha_2\beta_1}$ \\ \hline
\raisebox{-1.5ex}[0pt]{$P_s(IIB)$} & $G_{ij},\Phi(=0)$ & $\Psi^{\alpha_1}_{+i},\Psi^{\alpha_2}_{-i}$ & $(f^a_{11})^{\alpha_1\beta_1}, (f^a_{22})^{\alpha_2\beta_2}, (f_{12}^{\alpha_1\beta_2}-f_{21}^{\beta_2\alpha_1})$ \\
                             &  &  & $(f_{14}^{\alpha_1\beta_1}-f_{14}^{*\beta_1\alpha_1}), (f_{23}^{\alpha_2\beta_2}-f_{23}^{*\beta_2\alpha_2})$ \\ & & & $(f_{13}^{\alpha_1\beta_2}-f_{24}^{*\beta_2\alpha_1})$\\ \hline
\raisebox{0.0ex}[0pt]{Eff.} & $G^{eff}_{ij},\Phi(=0)$ & $(\Psi_{eff})^{\alpha_1}_i$  & $(f_{11}^{eff})^{\alpha_1\beta_1}, (f_{12}^{eff})^{\alpha_1\beta_2}, (f_{22}^{eff})^{\alpha_2\beta_2}$ \\ & & $(\Psi_{eff})^{\alpha_2}_i$ & $\bar f_{14}^{\alpha_1\beta_1}, \bar f_{23}^{\alpha_2\beta_2}, \bar f_{13}^{\alpha_1\beta_2}$ \\ \hline
\raisebox{0.0ex}[0pt]{Type I} & $G^I_{ij},\Phi^I(=0)$ & $(\Psi_I)^{\alpha_1}_{i},(\Psi_I)^{\alpha_2}_{i}$ & $(f_{11}^{a(I)})^{\alpha_1\beta_1}, (f_{22}^{a(I)})^{\alpha_2\beta_2}, (f_{12}^{(I)\alpha_1\beta_2}-f_{21}^{(I)\beta_2\alpha_1})$\\ & & & $(f_{14}^{(I)\alpha_1\beta_1}-f_{14}^{(I)\beta_1\alpha_1*}), (f_{23}^{(I)\alpha_2\beta_2}-f_{23}^{(I)\beta_2\alpha_2*})$\\ & & & $(f_{13}^{(I)\alpha_1\beta_2}-f_{24}^{(I)\beta_2\alpha_1*})$\\ \hline
\end{tabular}
\caption{Superstring theories and their background fields}
\end{table}

The last three rows contain the same sets of background fields.
Usual identification of type I theory with $P_s(IIB)$ preserves
only $\Omega$ even fields. We identify type I theory with
effective one and obtain improvement with squares of $\Omega$ odd
type IIB fields.

Consequently, while the $\Omega$ even fields appear in known
manner, the $\Omega$ odd fields, which are eliminated in standard
approach, here have two roles. They are source of noncommutativity
and their bilinear combinations are additional terms of effective
background. So, if naively, type I superstring theory, as $\Omega$
even one, can not recognize explicitly $\Omega$ odd fields of type
IIB theory, it can see them implicitly as $\Omega$ even
combinations in effective background.

\appendix

\section{Gamma matrices and spinors in $D=10$ and $d=6$ dimensions}
\setcounter{equation}{0}

The fermionic variables of the type IIB superstring theory,
$\theta^\alpha$ and $\bar\theta^\alpha$, are 10 dimensional
Majorana-Weyl spinors with 16 real components. In order to express
them in terms of general complex 32-component Dirac spinors, and
to investigate their relation with $D5$-brane spinors, we have to
introduce some representation of gamma matrices. We will use
notation and conventions introduced in Appendix B of the first
reference in \cite{jopol}. The gamma matrices in $D=10$ dimensions
are of the form $\Gamma^\mu=(\Gamma^i\, ,\Gamma^a)$
\begin{equation}\label{eq:Gama}
\Gamma^i=\gamma^i\otimes \left(
\begin{array}{cc}
\sigma_3 & 0\\
0 & -\sigma_3
\end{array}\right)\, ,\quad \Gamma^a={\bf{1}_8}\otimes \gamma^a\, ,
\end{equation}
where $\gamma^i\,(i=0,1,\dots,5)$ is $8\times 8$ representation of
gamma matrices in $d=6$ dimensions, while $\gamma^a\,(a=6,7,8,9)$
is $4\times 4$ representation of gamma matrices in $d=4$
dimensions. Matrices $\gamma^a$ are of the form
\begin{eqnarray}
&{}&\gamma^6=\left(
\begin{array}{cc}
-\sigma_1 & 0\\
0 & \sigma_1
\end{array}\right)
\, , \quad \gamma^7=\left(
\begin{array}{cc}
-\sigma_2 & 0\\
0 & \sigma_2
\end{array}\right)
\, ,\nonumber \\ &{}& \gamma^8=\left(
\begin{array}{cc}
0 & {\bf{1}_2}\\
{\bf{1}_2} & 0
\end{array}\right)
 \, ,\quad \gamma^9=\left(
\begin{array}{cc}
0 & -i {\bf{1}_2}\\
i {\bf{1}_2} & 0
\end{array}\right)\, ,
\end{eqnarray}
where the matrices $\sigma_i$ are Pauli $2\times 2$ matrices
\begin{equation}
\sigma_1=\left(
\begin{array}{cc}
0 & 1\\
1 & 0
\end{array}\right)\, ,\quad \sigma_2=\left(
\begin{array}{cc}
0 & -i\\
i & 0
\end{array}\right)\, ,\quad \sigma_3=\left(
\begin{array}{cc}
1 & 0\\
0 & -1
\end{array}\right)\, .
\end{equation}
For the corresponding $\gamma_{d+1}$ matrices in $D=10$ and $d=6$ dimensions we will use symbols $\Gamma$ and $\gamma$, respectively. According to the definition \cite{jopol}, we have
\begin{equation}
\Gamma\equiv \prod_{\mu=0}^9 \Gamma^\mu=\gamma\otimes \left(
\begin{array}{cc}
\sigma_3 & 0\\
0 & -\sigma_3
\end{array}\right)\, ,\quad \gamma\equiv -\prod_{i=0}^5 \gamma^i\, ,
\end{equation}
where $\Gamma$ and $\gamma$ are symmetric matrices, $\Gamma^T=\Gamma$ and $\gamma^T=\gamma$.

\subsection{Complex conjugation of gamma matrices}

The complex conjugation of gamma matrices can be described by operator $B_1$
\begin{equation}
B_1=\Gamma^3 \Gamma^5 \dots \Gamma^{d-1}\, ,
\end{equation}
which maps $\Gamma^\mu$ to $\Gamma^{\mu*}$
\begin{equation}
B_1 \Gamma^\mu B_1^{-1}=(-1)^{\frac{d-2}{2}} \Gamma^{\mu*}\, .
\end{equation}
In the case $D=10$ operator $B_1$ is of the form
\begin{equation}
B_1=\Gamma^3 \Gamma^5 \Gamma^7 \Gamma^9=b_1 \otimes \left(
\begin{array}{cc}
0 & i\sigma_2\\
i\sigma_2 & 0
\end{array}\right)
 \, ,
\end{equation}
where $b_1$ is the corresponding one for $d=6$
\begin{equation}
b_1=\gamma^3 \gamma^5\, .
\end{equation}
Matrix $B_1$ is symmetric, $B_1^T=B_1$, while the matrix $b_1$ is
antisymmetric, $b_1^T=-b_1$.

\subsection{Charge conjugation operator}

The transposed gamma matrices satisfy the same algebra as original ones.
There are similarity
transformations
which map from $\Gamma^\mu$ to $-\Gamma^{\mu\,T}$ and
$\gamma^i$ to $-\gamma^{i\,T}$
\begin{equation}
C \Gamma^\mu C^{-1}=-\Gamma^{\mu\,T}\, , \quad c\gamma^i
c^{-1}=-\gamma^{i\,T}\, ,
\end{equation}
described by charge conjugation operators
\begin{equation}\label{eq:opC}
C=B_1 \Gamma^0\, ,\quad c=b_1 \gamma^0\, .
\end{equation}

In $D=10$ dimensions operator $C$ is antisymmetric, $C^T=-C$, while the corresponding one in $d=6$ dimensions is symmetric, $c^T=c$.

\subsection{From ten dimensional to six dimensional spinors}

From $D=10$ Majorana and Weyl conditions
\begin{equation}
\theta^*=B_1 \theta\, ,\quad \Gamma \theta=\theta\, ,
\end{equation}
it follows that independent components of general 32 component Dirac spinor
\begin{equation}\label{eq:obliks}
\theta^\alpha=\left(
\begin{array}{c}
\theta_1\\
\theta_2\\
\theta_3\\
\theta_4
\end{array}\right)=\left(
\begin{array}{c}
\theta_1\\
\theta_2\\
b_1 \theta_2^*\\
-b_1 \theta_1^*
\end{array}\right)\, ,\quad (\alpha=1,2,\dots, 32)
\end{equation}
are two 8 component spinors $\theta_1^{\alpha_1}$ and $\theta_2^{\alpha_2}$ ($\alpha_1\, ,\alpha_2=1,2,\dots,8$) with constraints
\begin{equation}\label{eq:chir6}
\gamma \theta_1=\theta_1 \, ,\quad \gamma \theta_2=-\theta_2\, ,
\end{equation}
which we recognize as two opposite chirality Weyl spinors in $d=6$.
Similarly, from $\pi^{T\,*}=\pi^T B_1^T$ and $\pi^T \Gamma^T=\pi^T$, we obtain
\begin{equation}\label{eq:impuls6}
\pi_\alpha=\left(
\begin{array}{cccc}
\pi_1\, , & \pi_2\, , & -\pi_2^* b_1\, , & \pi_1^* b_1
\end{array}\right)\, ,
\end{equation}
with
\begin{equation}
\pi_1\gamma=\pi_1\, ,\quad \pi_2\gamma=-\pi_2\, .
\end{equation}
Consequently, we have
\begin{equation}\label{eq:relacija1}
\pi_\alpha \theta^\alpha=2\Re(\pi_{\alpha_1} \theta^{\alpha_1}+\pi_{\alpha_2}\theta^{\alpha_2})\, ,
\end{equation}
where symbol $\Re$ means real part of some complex number.

\section{Field strength $F^{\alpha\beta}$ in terms of antisymmetric tensors}
\setcounter{equation}{0}

The connection between two descriptions of R-R sector, field
strength $F^{\alpha\beta}$ and antisymmetric tensors $F_{(k)}$ can
be established \cite{grk}, where in short-hand notation $F_{(k)}$ denotes
$k$-rank antisymmetric tensor. It is known that
bispinor $F^{\alpha\beta}=S^\alpha (\Gamma^0 \tilde S)^\beta$,
made from same chirality spinors $S^\alpha$ and $\tilde
S^{\alpha}$ for type IIB theory, can be expand into complete set
of 10 dimensional antisymmetric gamma matrices
\begin{equation}\label{eq:RRpolje}
F^{\alpha\beta}=\sum_{k=0}^D
\frac{1}{k!}F_{(k)}\Gamma_{(k)}^{\alpha\beta}\, , \quad \left[
\Gamma_{(k)}^{\alpha\beta}=(C\Gamma^{[\mu_1\dots
\mu_k]})^{\alpha\beta}\right]
\end{equation}
where $C$ is charge conjugation operator defined in
(\ref{eq:opC}). Here
\begin{equation}
\Gamma^{[\mu_1 \mu_2\dots \mu_k]}\equiv \Gamma^{[\mu_1}\Gamma^{\mu_2}\dots \Gamma^{\mu_k]}\, ,
\end{equation}
is completely antisymmetrized product of gamma matrices.

The bispinor $F^{\alpha\beta}$ satisfy chirality
condition, $\Gamma F=-F\Gamma$, and, consequently, type IIB theory
contains only odd rank tensors $F_{(k)}$. Because of duality
relation, the independent tensors are $F_{(1)}$, $F_{(3)}$ and
self-dual part of $F_{(5)}$. Using mass-shell condition (massless
Dirac equation for $F^{\alpha\beta}$) these tensors can be solved
in terms of potentials $F_{(k)}=dA_{(k-1)}$, so that IIB theory
contains the potentials $A_{(0)}$, $A_{(2)}$ and $A_{(4)}$. The
number of independent components of $F^{\alpha\beta}$ is exactly
$256$, while the number of degrees of freedom is $64$.

The matrices $\Gamma_{(1)}$ and $\Gamma_{(5)}$ are symmetric in
spinor indices, while the matrix $\Gamma_{(3)}$ is
antisymmetric. So, the symmetric part of $F^{\alpha\beta}$,
$F_s^{\alpha\beta}=\frac{1}{2}(F^{\alpha\beta}+F^{\beta\alpha})$,
corresponds to the field strengths $F_{(1)}$ and $F_{(5)}$, and
antisymmetric part,
$F_a^{\alpha\beta}=\frac{1}{2}(F^{\alpha\beta}-F^{\beta\alpha})$,
corresponds to the field strength $F_{(3)}$.

Using the form of bispinor $F^{\alpha\beta}=S^{\alpha}(\Gamma^0
\tilde S)^\beta$, the form of spinors (\ref{eq:obliks}) and
expression for $\Gamma^0$ (\ref{eq:Gama}), we obtain
\begin{equation}\label{eq:jacinaF}
F=\left(
\begin{array}{cccc}
f_{11} & -f_{12} & f_{13}b_1 & f_{14}b_1\\
f_{21} & -f_{22} & f_{23}b_1 & f_{24}b_1\\
b_1 f_{24}^* & -b_1 f_{23}^* & b_1 f_{22}^* b_1 & b_1 f_{21}^* b_1\\
-b_1 f_{14}^* & b_1 f_{13}^* & -b_1 f_{12}^*b_1 & -b_1 f_{11}^* b_1
\end{array}\right)\, ,
\end{equation}
where
\begin{equation}
f_{11}=S_1 (\gamma^0 \tilde S_1)\, ,\quad f_{12}=S_1 (\gamma^0 \tilde S_2)\, ,\quad f_{21}=S_2 (\gamma^0 \tilde S_1)\, ,\quad f_{22}=S_2 (\gamma^0 \tilde S_2)\, ,
\end{equation}
\begin{equation}
f_{13}=S_1 (\gamma^0 \tilde S_2^*)\, ,\quad f_{14}=S_1 (\gamma^0 \tilde S_1^*)\, ,\quad f_{23}=S_2 (\gamma^0 \tilde S_2^*)\, ,\quad f_{24}=S_2 (\gamma^0 \tilde S_1^*)\, ,
\end{equation}
and $f_{11}$ corresponds to $f^{\alpha_1\beta_1}$, $f_{12}$ to $f^{\alpha_1\beta_2}$ etc.

The chirality condition $\Gamma F=-F \Gamma$ splits into eight conditions
\begin{equation}
\gamma f_{11}=-f_{11}\gamma\, ,\quad \gamma f_{12}=f_{12} \gamma\, ,\quad \gamma f_{21}=f_{21}\gamma\, ,\quad \gamma f_{22}=-f_{22}\gamma\, ,
\end{equation}
\begin{equation}
\gamma f_{13}=f_{13}\gamma\, ,\quad \gamma f_{14}=-f_{14}\gamma\, ,\quad \gamma f_{23}=-f_{23}\gamma \, ,\quad \gamma f_{24}=f_{24}\gamma\, .
\end{equation}
We are going to apply the same procedure as in $D=10$ and expand
$f_{11}$, $f_{12}$, $f_{21}$, $f_{22}$, $f_{13}$, $f_{14}$,
$f_{23}$ and $f_{24}$ into complete set of 6 dimensional
antisymmetric gamma matrices $\gamma_{(k)}\equiv
(c\gamma^{[i_1\dots i_k]})$. From the chirality conditions for
$f_{11}$, $f_{22}$, $f_{14}$ and $f_{23}$ it follows that they
contain odd rank tensors $f_{(1)}$ and self-dual part of
$f_{(3)}$, while from the chirality conditions for $f_{12}$,
$f_{21}$, $f_{13}$ and $f_{24}$ it follows that they contain even
rank tensors $f_{(0)}$ and $f_{(2)}$. Using mass-shell condition
these tensors can be expressed in terms of potentials as
$f_{(k)}=d a_{(k-1)}$. Consequently, $f_{11}$, $f_{22}$, $f_{14}$
and $f_{23}$ contain potentials $a_{(0)}$ and self-dual part of
$a_{(2)}$, while $f_{12}$, $f_{21}$, $f_{13}$ and $f_{24}$ contain
potential $a_{(1)}$ ($a_{(-1)}$ is not a physical degree of
freedom). The number of independent components of $f_{11}$,
$f_{22}$, $f_{14}$, $f_{23}$, $f_{12}$, $f_{21}$, $f_{13}$ and
$f_{24}$ is $128$, while the number of degrees of freedom is $32$.

In analogy with the case $D=10$, in $d=6$ dimensions, symmetric
parts of $f_{11}$, $f_{22}$, $f_{14}$ and $f_{23}$ correspond to
the field strength $f_{(3)}$, while their antisymmetric parts
correspond to the field strength $f_{(1)}$.

Using the expressions (\ref{eq:impuls6}) and (\ref{eq:jacinaF}), we have
\begin{eqnarray}\label{eq:pFp}
&\pi&_\alpha F^{\alpha\beta}\bar\pi_\beta= \\&=&2\Re (\pi_{\alpha_1} f_{11}^{\alpha_1\beta_1}\bar\pi_{\beta_1}+\pi_{\alpha_1} f_{14}^{\alpha_1\beta_1}\bar\pi_{\beta_1}^*-\pi_{\alpha_2} f_{22}^{\alpha_2\beta_2}\bar\pi_{\beta_2}-\pi_{\alpha_2} f_{23}^{\alpha_2\beta_2}\bar\pi_{\beta_2}^*)\nonumber \\&+&2\Re(\pi_{\alpha_2} f_{21}^{\alpha_2\beta_1}\bar\pi_{\beta_1}-\pi_{\alpha_1} f_{12}^{\alpha_1\beta_2}\bar\pi_{\beta_2}+\pi_{\alpha_2} f_{24}^{\alpha_2\beta_1}\bar\pi_{\beta_1}^*-\pi_{\alpha_1} f_{13}^{\alpha_1\beta_2}\bar\pi_{\beta_2}^*)\, .\nonumber
\end{eqnarray}

\section{Complex coordinates and their canonically conjugated momenta}
\setcounter{equation}{0}

Lagrangian contains complex coordinates and momenta. It is important to determine the basic Poisson structure between coordinates and momenta.

Let us start with one term from the Lagrangian (\ref{eq:SB1})
\begin{equation}\label{eq:term}
2\Re\left[-\pi_{\alpha_1}(\partial_\tau-\partial_\sigma) \theta^{\alpha_1}\right]\, .
\end{equation}
Decomposing $\theta^{\alpha_1}$ and $\pi_{\alpha_1}$ in real and imaginary parts
\begin{equation}
\theta^{\alpha_1}=\theta^{\alpha_1}_1+i\theta^{\alpha_1}_2\, ,\quad \pi_{\alpha_1}=\pi_{\alpha_1}^1+i\pi_{\alpha_1}^2\, ,
\end{equation}
expression (\ref{eq:term}) gets the form
\begin{equation}
-2\pi_{\alpha_1}^1 (\partial_\tau-\partial_\sigma)\theta^{\alpha_1}_1+2\pi_{\alpha_1}^2 (\partial_\tau-\partial_\sigma)\theta^{\alpha_1}_2\, .
\end{equation}
The canonically conjugated momentum for $\theta^{\alpha_1}_1$ is $2\pi_{\alpha_1}^1$, while for $\theta^{\alpha_1}_2$ is $-2\pi_{\alpha_1}^2$, which means that all nonzero Poisson brackets are
\begin{equation}
\left\lbrace \theta^{\alpha_1}_1(\sigma)\, ,\pi_{\beta_1}^1(\bar\sigma)\right\rbrace=-\frac{1}{2}\delta^{\alpha_1}{}_{\beta_1}\delta(\sigma-\bar\sigma)\, ,\quad \left\lbrace \theta^{\alpha_1}_2(\sigma)\, ,\pi_{\beta_1}^2(\bar\sigma)\right\rbrace=\frac{1}{2}\delta^{\alpha_1}{}_{\beta_1}\delta(\sigma-\bar\sigma)\, .
\end{equation}
Using these relations we easily obtain Poisson brackets of the complex variables
\begin{equation}
\left\lbrace \theta^{\alpha_1}(\sigma)\, ,\pi_{\beta_1}(\bar\sigma)\right\rbrace=-\delta^{\alpha_1}{}_{\beta_1}\delta(\sigma-\bar\sigma) \, ,\quad \left\lbrace \theta^{\alpha_1}(\sigma)\, ,\pi^*_{\beta_1}(\bar\sigma)\right\rbrace=0\, ,\quad \left\lbrace \theta^{*\alpha_1}(\sigma)\, ,\pi_{\beta_1}(\bar\sigma)\right\rbrace=0\, ,
\end{equation}
where ${}^*$ means complex conjugation. This means that
$\theta^{\alpha_1}$ and $\pi_{\alpha_1}$ are canonically
conjugated complex variables, while $\theta^{\alpha_1}$ and
$\pi_{\beta_1}^*$ are canonically independent. The same procedure
can be repeated for all other terms in Lagrangian.

Similarly, in definition of canonical Hamiltonian we have
\begin{equation}
\dot \theta_1^{\alpha_1} 2\pi^1_{\alpha_1}-\dot \theta^{\alpha_2}_2 2\pi^2_{\alpha_2}=2\Re(\dot \theta^{\alpha_1}\pi_{\alpha_1})=2\Re(\dot \theta^{*\alpha_1}\pi^*_{\alpha_1})\, .
\end{equation}

\section{Consistency procedure for fermionic constraints - explicit expressions}
\setcounter{equation}{0}

In order to investigate the consistency of the fermionic
constraints (\ref{eq:bcf1}) and (\ref{eq:bcf2}), we have to apply consistency procedure to the variables
$$A^{(0)}=(\theta^{\alpha_1}, \theta^{\alpha_2}, \bar\theta^{\alpha_1}, \bar\theta^{\alpha_2}, \pi_{\alpha_1}, \pi_{\alpha_2}, \bar\pi_{\alpha_1}, \bar\pi_{\alpha_2})\, .$$
If we define the recurrent equation
\begin{equation}
A^{(n+1)}\equiv\{H_c\, ,A^{(n)}\}\, ,\quad (n\geq 0)
\end{equation}
we obtain
\begin{equation}\label{eq:koordinate1}
(\Theta^{\alpha_1})^{(2n)} =\partial_\sigma^{(2n)} \theta^{\alpha_1} \, ,\quad (\Theta^{\alpha_2})^{(2n)} =\partial_\sigma^{(2n)} \theta^{\alpha_2}\, ,
\end{equation}
\begin{eqnarray}
(\Theta^{\alpha_1})^{(2n+1)}
&=&-\partial_{\sigma}^{(2n+1)}\theta^{\alpha_1}-\Psi^{\alpha_1}_i
\partial_{\sigma}^{(2n+1)}x^i-\frac{1}{2\kappa}f^{\alpha_1\beta_{1}}_{11}\partial_{\sigma}^{(2n)}\bar\pi_{\beta_1}-\frac{1}{2\kappa}f^{\alpha_1\beta_{1}}_{14}\partial_{\sigma}^{(2n)}\bar\pi_{\beta_1}^*\nonumber \\&+&\frac{1}{2\kappa}f_{12}^{\alpha_1\beta_2}\partial_{\sigma}^{(2n)}\bar\pi_{\beta_2}+\frac{1}{2\kappa}f_{13}^{\alpha_1\beta_2}\partial_{\sigma}^{(2n)} \bar\pi_{\beta_2}^*+\frac{1}{2\kappa}G^{ij}\partial_{\sigma}^{(2n)}(I_{+
i}+I_{- i})\Psi^{\alpha_1}_j\, , \nonumber \\
(\Theta^{\alpha_2})^{(2n+1)}
&=&-\partial_{\sigma}^{(2n+1)}\theta^{\alpha_2}-\Psi^{\alpha_2}_i
\partial_{\sigma}^{(2n+1)}x^i+\frac{1}{2\kappa}f^{\alpha_2\beta_{2}}_{22}\partial_{\sigma}^{(2n)}\bar\pi_{\beta_{2}}+\frac{1}{2\kappa}f^{\alpha_2\beta_{2}}_{23}\partial_{\sigma}^{(2n)}\bar\pi_{\beta_{2}}^*\nonumber \\&-&\frac{1}{2\kappa}f_{21}^{\alpha_2\beta_1}\partial_{\sigma}^{(2n)}\bar\pi_{\beta_1}-\frac{1}{2\kappa}f_{24}^{\alpha_2\beta_1}\partial_{\sigma}^{(2n)}\bar\pi_{\beta_1}^*+\frac{1}{2\kappa}G^{ij}\partial_{\sigma}^{(2n)}(I_{+
i}+I_{- i})\Psi^{\alpha_2}_j
,
\end{eqnarray}
\begin{equation}
(\bar\Theta^{\alpha_1})^{(2n)}=\partial_\sigma^{(2n)}\bar\theta^{\alpha_1}\,
,\quad (\bar\Theta^{\alpha_2})^{(2n)}=\partial_\sigma^{(2n)}\bar\theta^{\alpha_2}\, ,
\end{equation}
\begin{eqnarray}
(\bar\Theta^{\alpha_1})^{(2n+1)}&=&
\partial_{\sigma}^{(2n+1)}\bar\theta^{\alpha_1}+\bar\Psi^{\alpha_1}_i \partial_{\sigma}^{(2n+1)}x^i+\frac{1}{2\kappa}\partial_\sigma^{(2n)}\pi_{\beta_{1}}
f^{\beta_{1}\alpha_1}_{11}+\frac{1}{2\kappa}\partial_\sigma^{(2n)}\pi_{\beta_{1}}^*
f^{*\beta_{1}\alpha_1}_{14}\nonumber \\
&+&\frac{1}{2\kappa}f_{21}^{\beta_2\alpha_1}\partial_{\sigma}^{(2n)}\pi_{\beta_2}+\frac{1}{2\kappa}f_{24}^{*\beta_2\alpha_1}\pi_{\beta_2}^*+\frac{1}{2\kappa}G^{ij}\partial_\sigma^{(2n)} (I_{+ i}+I_{-
i})\bar\Psi^{\alpha_1}_{j}\, ,\nonumber \\
(\bar\Theta^{\alpha_2})^{(2n+1)}&=&
\partial_{\sigma}^{(2n+1)}\bar\theta^{\alpha_2}+\bar\Psi^{\alpha_2}_i \partial_{\sigma}^{(2n+1)}x^i -\frac{1}{2\kappa}\partial_\sigma^{(2n)}\pi_{\beta_{2}}
f^{\beta_{2}\alpha_2}_{22}-\frac{1}{2\kappa}\partial_\sigma^{(2n)}\pi^*_{\beta_{2}}
f^{*\beta_{2}\alpha_2}_{23}\nonumber \\
&-&\frac{1}{2\kappa}f_{12}^{\beta_1\alpha_2}\partial_\sigma^{(2n)}\pi_{\beta_1}-\frac{1}{2\kappa}f_{13}^{*\beta_1\alpha_2}\partial_\sigma^{(2n)}\pi_{\beta_1}^*+\frac{1}{2\kappa}G^{ij}\partial_\sigma^{(2n)} (I_{+ i}+I_{-
i})\bar\Psi^{\alpha_2}_{j} \label{eq:koordinate2}
\end{eqnarray}
\begin{eqnarray}\label{eq:impulsi}
(\Pi_{\alpha_1})^{(n)} &=&  (-1)^n \partial_\sigma^{(n)}\pi_{\alpha_1}\,
,\quad (\Pi_{\alpha_2})^{(n)} =  (-1)^n \partial_\sigma^{(n)}\pi_{\alpha_2}\, ,\nonumber \\ (\bar\Pi_{\alpha_1})^{(n)} &=&
\partial_\sigma^{(n)}\bar\pi_{\alpha_1}\, ,\quad (\bar\Pi_{\alpha_2})^{(n)} =
\partial_\sigma^{(n)}\bar\pi_{\alpha_2}\, .
\end{eqnarray}

Defining the function
\begin{equation}
\mathcal A(\sigma)\equiv\sum_{n=0}^\infty
\frac{\sigma^n}{n!}A^{(n)}(\sigma=0)\, ,
\end{equation}
we introduce compact $\sigma$ dependent expressions for the corresponding variables after consistency procedure
\begin{eqnarray}
&{}&\Theta^{\alpha_1}(\sigma)=\theta^{\alpha_1}(-\sigma)-\Psi^{\alpha_1}_i
\tilde q^i(\sigma)-\frac{1}{2\kappa}
f^{\alpha_1\beta_{1}}_{11}\int_0^\sigma d\sigma_1 P_s
\bar\pi_{\beta_{1}}-\frac{1}{2\kappa}
f^{\alpha_1\beta_{1}}_{14}\int_0^\sigma d\sigma_1 P_s
\bar\pi_{\beta_{1}}^*\nonumber \\&+&\frac{1}{2\kappa}f_{12}^{\alpha_1\beta_2}\int_0^\sigma d\sigma_1 P_s \bar\pi_{\beta_2}+\frac{1}{2\kappa}f_{13}^{\alpha_1\beta_2}\int_0^\sigma d\sigma_1 P_s \bar\pi_{\beta_2}^*+\frac{1}{2\kappa}G^{ij}
\Psi^{\alpha_1}_i   \int_0^\sigma d\sigma_1 P_s (I_{+ j}+I_{- j})
,\nonumber \\
&{}&\Theta^{\alpha_2}(\sigma)=\theta^{\alpha_2}(-\sigma)-\Psi^{\alpha_2}_i
\tilde q^i(\sigma)+\frac{1}{2\kappa}
f^{\alpha_2\beta_{2}}_{22}\int_0^\sigma d\sigma_1 P_s
\bar\pi_{\beta_{2}}+\frac{1}{2\kappa}
f^{\alpha_2\beta_{2}}_{23}\int_0^\sigma d\sigma_1 P_s
\bar\pi_{\beta_{2}}^*\nonumber \\&-&\frac{1}{2\kappa}f_{21}^{\alpha_2\beta_1}\int_0^\sigma d\sigma_1 P_s \bar\pi_{\beta_1}-\frac{1}{2\kappa}f_{24}^{\alpha_2\beta_1}\int_0^\sigma d\sigma_1 P_s \bar\pi_{\beta_1}^*\nonumber \\ &+&\frac{1}{2\kappa}G^{ij}
\Psi^{\alpha_2}_i   \int_0^\sigma d\sigma_1 P_s (I_{+ j}+I_{- j})
\, ,\label{eq:Fialfa}
\end{eqnarray}
\begin{eqnarray}
&{}&\bar\Theta^{\alpha_1}(\sigma)=\bar\theta^{\alpha_1}(\sigma)+\bar\Psi^{\alpha_1}_i
\tilde q^i(\sigma)+\frac{1}{2\kappa}f^{\beta_{1}\alpha_1}_{11}
\int_0^\sigma d\sigma_1 P_s
\pi_{\beta_{1}}+\frac{1}{2\kappa}f^{*\beta_{1}\alpha_1}_{14}
\int_0^\sigma d\sigma_1
P_s \pi_{\beta_{1}}^* \,\,\nonumber \\
&+&\frac{1}{2\kappa}f_{21}^{\beta_2\alpha_1}\int_0^\sigma d\sigma_1 P_s \pi_{\beta_2}+\frac{1}{2\kappa}f_{24}^{*\beta_2\alpha_1}\int_0^\sigma d\sigma_1 P_s \pi_{\beta_2}^*+\frac{1}{2\kappa}G^{ij} \bar\Psi^{\alpha_1}_i \int_0^\sigma
d\sigma_1 P_s (I_{+ j}+I_{- j})\, ,\nonumber \\
&{}&\bar\Theta^{\alpha_2}(\sigma)=\bar\theta^{\alpha_2}(\sigma)+\bar\Psi^{\alpha_2}_i
\tilde q^i(\sigma)-\frac{1}{2\kappa}
f^{\beta_{2}\alpha_2}_{22} \int_0^\sigma d\sigma_1 P_s
\pi_{\beta_{2}}-\frac{1}{2\kappa} f^{*\beta_{2}\alpha_2}_{23}
\int_0^\sigma d\sigma_1 P_s \pi^*_{\beta_{2}}\nonumber \\
&-&\frac{1}{2\kappa}f_{12}^{\beta_1\alpha_2}\int_0^\sigma d\sigma_1 P_s \pi_{\beta_1}-\frac{1}{2\kappa}f_{13}^{*\beta_1\alpha_2}\int_0^\sigma d\sigma_1 P_s \pi_{\beta_1}^*\nonumber \\&+&\frac{1}{2\kappa}G^{ij} \bar\Psi^{\alpha_2}_i \int_0^\sigma
d\sigma_1 P_s (I_{+ j}+I_{- j})\, ,\label{eq:barFialfa}
\end{eqnarray}

\begin{eqnarray}\label{eq:Pialfa}
\Pi_{\alpha_1}(\sigma)&=&\pi_{\alpha_1}(-\sigma)\, ,\quad
\Pi_{\alpha_2}(\sigma)=\pi_{\alpha_2}(-\sigma)\, ,\nonumber \\
\bar\Pi_{\alpha_1}(\sigma)&=&\bar\pi_{\alpha_1}(\sigma)\, ,\quad
\bar\Pi_{\alpha_2}(\sigma)=\bar\pi_{\alpha_2}(\sigma)\, .
\end{eqnarray}


\begin{thebibliography}{99}

\bibitem{BNBSPLB} B. ~Nikoli\'c and B. ~Sazdovi\'c, {\it Phys. Lett.} {\bf B666} (2008) 400.
\bibitem{jopol} J. Polchinski, {\it String theory - Volume II}, Cambridge University Press,
1998; K. Becker, M. Becker and J. H. Schwarz, {\it String Theory
and M-Theory - A Modern Introduction}, Cambridge University Press,
2007.


\bibitem{berko} N. ~Berkovits, hep-th/0209059; P. ~A. ~Grassi, G. ~Policastro and P. ~van ~Nieuwenhuizen, {\it JHEP} {\bf 10} (2002) 054; P. ~A. ~Grassi, G. ~Policastro and P. ~van ~Nieuwenhuizen, {\it JHEP} {\bf 11} (2002) 004; P. ~A. ~Grassi, G. ~Policastro and P. ~van ~Nieuwenhuizen, {\it Adv. Theor. Math. Phys.} {\bf 7} (2003) 499; P. ~A. ~Grassi, G. ~Policastro and P. ~van ~Nieuwenhuizen, {\it Phys. Lett.} {\bf B553} (2003) 96.



\bibitem{susyNC} J. ~de ~Boer, P. ~A. ~Grassi and P. ~van ~Nieuwenhuizen, {\it Phys. Lett.} {\bf B574} (2003) 98.
\bibitem{NPBref} N. ~Berkovits and P. Howe, {\it Nucl. Phys.} {\bf B635} (2002) 75.
\bibitem{duf} M. ~J. ~Duff, Ramzi ~R. ~Khuri and J. ~X. ~Lu, {\it Phys. Rept.} {\bf 259} (1995) 213.
\bibitem{grk} E. Kiritsis, \textit{Introduction to Superstring Theory}, Leuven University Press, 1998, hep-th/9709062.
\bibitem{SW} N. Seiberg and E. Witten, {\it JHEP} {\bf 09} (1999) 032.
\bibitem{JT} T. H. Buscher, {\it Phys. Lett.} {\bf B201} (1988) 466; A. Giveon, M. Porrati and E. Rabinovici, {\it Phys.Rept.} {\bf 244} (1994) 77; R. Benichou, G. Policastro and J. Troost, {\it Phys. Lett.} {\bf B661} (2008) 192.

\bibitem{radepjc} B. ~Nikoli\'c and B. ~Sazdovi\'c, arXiv:0711.4463 (accepted for publication in Advances in Theoretical
and Mathematical Physics, Issue February 2010 - in press).

\bibitem{BNBS} B. ~Sazdovi\'c, {\it Eur. Phys. J} {\bf C44} (2005) 599; B. ~Nikoli\'c and B. ~Sazdovi\'c, {\it Phys. Rev.} {\bf D74} (2006) 045024.

\bibitem{BNBS2} B. ~Nikoli\'c and B. ~Sazdovi\'c, {\it Phys. Rev.} {\bf D75} (2007) 085011.





















\end{thebibliography}
\end{document}